**DEPARTMENT OF ELECTRICAL AND ELECTRONIC ENGINEERING**

# CONTROL SCHEMES FOR DISTRIBUTION GRIDS WITH MASS DISTRIBUTED GENERATION

**AUTHOR: Umair Shahzad**

**SUPERVISOR: Prof Greg Asher**

**DATE: September 2012**

Project thesis submitted in part fulfilment of the requirements for the degree of

Master of Science in Electrical Engineering, The University of Nottingham.

# Abstract

This project discusses the control schemes for distribution grids with a large amount of wind penetration. Microgrids are constantly gaining popularity, especially in the countries, where there is energy crisis. Various systems, including synchronous generators, grid and loads, have been investigated in this project. Major focus is placed on active and reactive power sharing. Droop control for multiple synchronous generators is explored. The phenomenon of load transients has also been reviewed and associated simulations have been carried out on SimPower Systems. Constant wind power has been introduced and behaviour of the electrical system is observed. Behaviour of the system under variable wind has also been analysed. Moreover, recent development projects and previous works, regarding microgrids and distributed generation, have been discussed.

# **Acknowledgements**

Firstly, I would like to thank my supervisor, Professor Greg Asher, for his invaluable help and constructive advice, for his patience, enthusiasm and immense technical knowledge. I would like to express my heartiest gratitude to him whose wide range of expertise and knowledge has helped me make this project a success.

My special thanks are also extended to Dr. Serhiy Bozhko, Mr. Sung Oe and Mr. Shuai Shao for their numerous guidelines during my project.

Last, but not the least, I would like to express my deepest love and appreciation to my family, especially, my parents and grandparents, for their never-ending love, constant support and words of comfort.

# List of Abbreviations

**DG:** Distributed Generation

**MG:** Microgrid

**LV:** Low Voltage

**MV:** Medium Voltage

**CHP:** Combined Heat and Power

**DC:** Direct Current

**AC:** Alternating Current

**PV:** Photovoltaic

**SD:** Separation Device

**VSI:** Voltage Source Inverter

**SCR:** Silicon Controlled Rectifier

**PCC:** Point of Common Coupling

**RMS:** Root Mean Square

**FACTS:** Flexible Alternating Current Transmission System

**RPM:** Revolutions per Minute

**VAR**: Volt-Ampere Reactive

**SG:** Synchronous Generator

**PI**: Proportional Integral

**UPS:** Uninterruptible Power Supply

# Glossary

**Microgrid:** It is a simple network which provides energy/power into the system when the main grid fails. In other words, it an alternative means for delivering power to the consumers. It is commonly abbreviated as MG.

**Distributed Generation**: It is commonly abbreviated as DG. This is the main part of a microgrid. It has various types like wind, solar, fuel cell etc. They generate power and feed it into the main power system in case the main grid collapses.

**Droop Control**: It is basically a mechanism for controlling the active and reactive power sharing of multiple generators. These powers are controlled according to the slopes (droop coefficients) of the active power-frequency and reactive-power graphs.

**Master-Slave:** It is a very common term used in microgrids. It simply means that one system sets the frequency (and hence speed) of the other. The second system follows the first one. In other words, one system acts as the "Master" and the other one as the "Slave".

**Grid-Connected Mode**: This is the mode in which the grid is connected to the system and is constantly supplying power.

**Grid-Disconnected Mode:** It is also called island, stand-alone or emergency mode. In this mode, grid disconnects from the system and microgrid comes into action.

**Synchronisation**: It is a process which is carried out to connect multiple generators to an infinite bus bar. For this, the rms line voltages, phase angles, phase sequences and frequencies of the generators must be same.

**CHP:** It is an acronym for combined heat and power systems. Such systems, besides delivering power, produce some waste heat from the gases. This heat can be utilized and this improves the efficiency of the network.

**Micro-source:** It is a source of distributed generation which is used in micro-grids. It can be wind, solar, fuel cells etc.

**Critical Loads**: These are the loads which are dependent on microgrid. In other words, the loads to which the micro-source supplies power are called critical loads. They are also called sensitive and non-traditional loads.

**Non-Critical Loads**: These are the loads which are cut off from the system when microgrid starts delivering power. They are also known as non-sensitive and traditional loads.

# Table of Contents

# CHAPTER 1

# <u>INTRODUCTION</u>

This chapter will introduce the background of the project, the main aims and the objectives and how they will be achieved. Thesis structure and layout is also presented.

## 1.1 Introduction

As the power demand is increasing day by day, the importance of producing more energy cannot be neglected. In this regard, micro-grids, also known as distributed grids, are of utmost importance. They are a good means of providing energy to the network in case the main grid fails. This project deals with the control schemes for distribution grids with mass distributed generation. It is likely that in near future, there will be more reliance on distributed generation, especially at low voltage (LV) and medium voltage (MV) levels. Such systems will increase the overall stability and reliability of the power network and at the same time, will provide more efficient performance. In simple words, if there is any fault on main grid and it is tripped, the micro-grid does not disrupt power flow and continues to supply power to the consumers.

## 1.2 Aims and Objectives

The major aim of the project will be to develop the control methods for micro-grids under high levels of distributed generation which, in case of this project, is wind power. To achieve this aim, the software called SimPower Systems will be used. It is basically a part of MATLAB software. Several basic systems will be built and then, the final modeling will be shown. The concept of master-slave and droop-based control of multiple synchronous generators will be investigated. Load transients will also be observed. Finally, the wind power will be introduced in the system and active and reactive power sharing will be explored.

At each stage of the project, behavior of different parameters like voltages, currents, active power, reactive power, speed etc. will be observed and discussed.

Wind power will be used as distributed generation source. Literature [7] has proved that wind power is the first preferred choice for micro-source. Previous research results [13] have shown that distributed generation (DG) could have significant impact on the stability of the transmission system, especially at large penetration levels, where penetration is DG power as a percentage of total load power.

## 1.3 Thesis Structure and Layout

The first chapter will deal with the general introduction about the project. Major aims and objectives will also be mentioned. The second chapter presents research work carried out in the past regarding microgrids and their control schemes. The theoretical background and recent development projects, regarding microgrids and distributed generation, are also covered in this chapter. The third chapter will give some insight into simulations of two major networks. The first network will consist of the main grid, a synchronous generator and various loads. The second one will include two synchronous generators and loads. Behaviour of various electrical quantities will be observed. Active and reactive powers for generators and loads will also be explored in this chapter. The fourth chapter will look into the droop control of these generators and how active and reactive power is shared between them according to the droop coefficients. The fifth chapter will give some idea about the behaviour of the system when wind power (source of distributed generation) is introduced into the system. The sixth (and the last) chapter will conclude the project with insight into guidelines for future work regarding microgrids.

# CHAPTER 2

# LITERATURE REVIEW AND BACKGROUND INFORMATION

This chapter reviews some research work and recent development projects regarding microgrids and distributed generation. Detailed information, related to various issues and factors of microgrid and distributed generation, is also elaborated.

## 2.1 Literature Review and Background Information

Microgrids provide more efficiency and reliability in the power network. Efficiency can be considerably enhanced using CHP (Combined Heat and Power) techniques. Piagi and Lasseter [1] have found that during control mode of micro-grid, there must be some energy balance between power supply and demand. This is possible by dispatching generators and/or loads. They [1] have proposed a micro-source control scheme for facilitating seamless mode transfer between grid and micro-grid. It means in case of a fault, power can be rapidly detached from grid and shifted to micro-grid. The authors [1] have used power-frequency droop to implement this. Similar idea will be used in the present project to investigate power sharing between the two synchronous generators. They have carried out a case study on University of Wisconsin microgrid to demonstrate the practical implications of island and grid connected modes. [1]

S. Krishnamurthy et al. [2] have studied the operation of diesel engine fed synchronous generator sets as distributed generation sources. They found that controlling the reactive power-voltage droop characteristics is very essential; otherwise large reactive currents would flow in the synchronous generators.

Ling Su et al. [3] carried out studies on the micro-grid control schemes using two micro-turbines as micro-sources. They slightly changed the way the droop parameters are designed. They enforced a limit on the active power output so that when power is at the maximum value, frequency is at its minimum value and vice versa. Yong Xue et al. [4] designed a control scheme to control the active power in a distributed generation unit in grid-connected mode. Similarly,

Basak P et al. [9] have discussed the control techniques for island mode of micro-grids. One major conclusion drawn from their work is that when micro-grid is operating in island mode, all non-critical (traditional) loads are eliminated automatically. Zhang Jie et al. [10] have used the inverter control technique for micro-grid control. This is the most common control technique. They also studied the transition between grid-connected and grid-disconnected mode. They have used Voltage Source Inverter (VSI) as micro-source. Wei Huang et al. [11] have studied seamless transfer between grid-connected and disconnected mode using SCR trigger control.

Jia Yaoqin et al. [5] have carried out research on standalone micro-grid at low voltage. They used parallel inverters as sources of distributed generation. They have used an improved method of droop control to stabilize the output voltage which results in small reactive currents.

Rowe C et al. [8] have designed the power-frequency droop using a voltage PI controller in the direct axis of rotating reference frame.

Huang Jiayi et al. [12] have discussed the current situation and projects about micro-grids in Europe and Japan. In Portugal, a micro-grid has been developed which is supplied by a LV feeder from a distributed power station of 200 KVA. Similarly, in Japan, an organization has started three research projects in this area. In one of these, they have used fuel cells as micro-source.

A vital part of microgrid operation consists of energy management and controlling scheme in islanded mode. Literature [14] has put forward the technique of super capacitor and battery combination for this. Flywheel may be used in some cases.

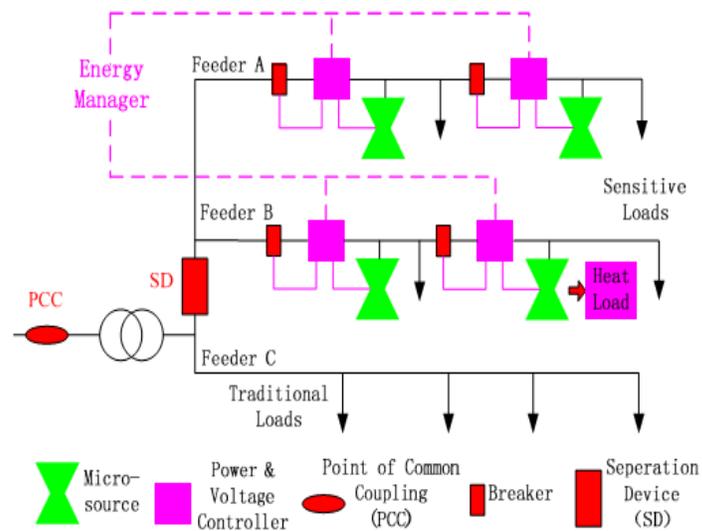

**Figure 2.1 Typical Distributed Grid System** [11]

The separation device (marked as SD) in Figure 2.1, taken from [11], is basically a static switch. When it is open, we have what we call as the island mode or stand-alone mode. Under this mode, all traditional loads (non-critical) are eliminated and the micro-source powers the sensitive (critical) loads only. The details about microgrid are presented in Section 2.3.

## 2.2 Distributed Generation

### 2.2.1 Definitions

According to IEEE, distributed generation is the "generation of electricity by the facilities that are sufficiently smaller than central generating plants so as to allow interconnection at nearly any point in a power system". According to Dondi et al., distributed generation is the generation of electrical power using a small source which is not part of the large central power system and which is located in close vicinity of the load. Similarly, according to Ackermann et al., distributed generation is the generation of electrical power from the source which is directly connected to the distribution network. It is also known as dispersed or embedded generation in USA. In some European countries, it is called decentralized generation. It must be noted that distributed generation should not be confused with renewable energy because distributed generation may include renewable or non-renewable technologies as will be explained later on. Distributed generation is usually abbreviated as DG in literature. [17], [18].

According to Arthur D.Little, "Distributed generation is the integrated or stand- alone use of small, modular electricity generation resources by utilities, utility customers and third parties in applications that benefit the electric system, specific end-user customers or both." [19].

## 2.2.2 Why to use Distributed Generation?

There are many factors which contribute in the attraction towards distributed generation. Firstly, it gives a vital opportunity to different players in the electricity market to provide electric power according to customer requirements. In other words, there is more flexibility in providing electricity. Secondly, they can provide electric power to local loads, hence, saving the costs of grid connections to remote and geographically challenging areas. They can also provide grid support, for instance, stabilizing a dropping frequency (due to some fault or excess load current). Distributed generation is environmentally beneficial too. It is because most of the distributed generation technologies include renewable sources like wind, solar etc. They can produce combined heat and power when CHPs are used. This improves the efficiency significantly. [18]

The trend towards using distributed generation has increased significantly in the last few years. Figure 2.2, taken from [21], shows the distributed generation growth annually over the past few years.

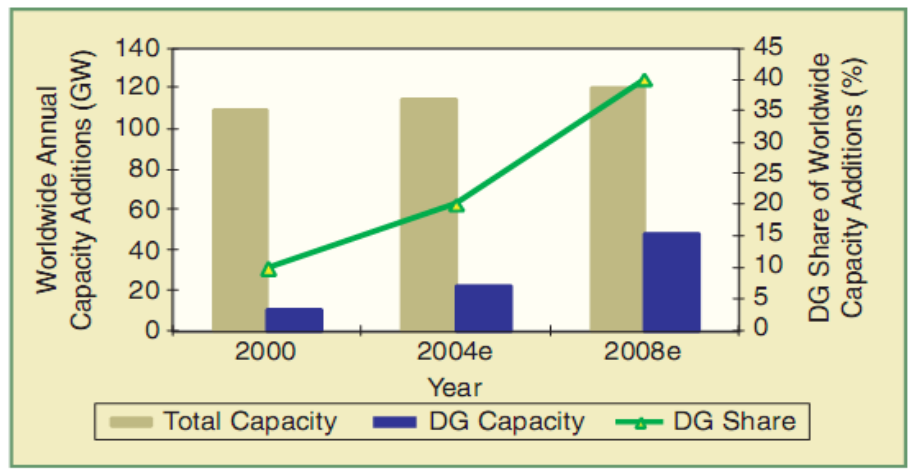

**Figure 2.2 Worldwide annual trends in distributed generation** [21]

## 2.2.3 Types of Distributed Generation

There are many types of distributed generation [17], [20]. Some of them are explained below:

**Reciprocating Engines**: This technology is one of the oldest distributed generation technologies. The engines commonly use natural gas or diesel as their fuel source. The design of the engine is particularly important for increased efficiency and reduction in the emission nitrous oxides.

**Micro-turbines**: This is one the popular technologies in distributed generation. In their simplest form, they consist of a small turbine, compressor and a generator. They are often packed together in the form of units. Their output typically varies from 30 KWs to 200 KWs. Their main advantage is that they give out very low gaseous emissions and are generally environmentally friendly. They require very little maintenance. But the major obstacle in their use is their comparatively high cost and reduced efficiency. Recuperators can significantly improve the efficiency. These are basically air to air heat exchangers which use the heat from the waste gases to pre-heat the inlet for air .Common types are simple cycle and combined cycle turbines.

**Fuel Cells**: They basically convert chemical energy into electrical energy by employing some electrochemical processes. They require hydrogen for their operation but they use propane or natural gas as fuel source. It is because using hydrogen directly for their operation is not feasible. This technology just like micro-turbines has low emissions but faces the problem of high costs and reliability issues. They do not have any rotating parts; hence, they require minimum maintenance. They have different types depending on the electrolytic material employed. Some common examples include Phosphoric Acid Fuel Cell (PAFC), alkaline fuel cells and Proton Exchange Membrane Fuel Cell (PEMFC).

**Photo-voltaics**: They are one of the popular technologies used in distributed generation. They are also known as solar panels. They have very low emission rates and require very less repair and maintenance.  They are in the form of small modules and hence, can be placed where there is maximum amount of sunshine. Relatively high cost of purchase and installation is the main issue which should be resolved in order to make them commercially more viable. They are mostly used in remote locations where there is no feasibility of grid connections. They are also common in space applications to provide power to satellites.

**Wind Power**: Wind turbines offer a relatively cheap method to provide electricity. They have low $CO_2$ emissions and low pollution effects. The major hurdle in their use is that wind is unpredictable and hence, we are not sure of the power output. Secondly, wind turbines require a lot of maintenance regarding gearbox and rotor. They are also hazardous for birds and bats. In order to counter-act the unpredictable nature of wind velocity, battery storage systems are being incorporated into the networks to continue the supply of electricity when the turbine blades are not rotating. Voltage regulators help these batteries to charge and store energy.

Further, the distributed generation sources can also be categorized as renewable and non renewable technologies.

Renewable technologies consist of solar, wind, geothermal etc. while non renewable technologies include fuel cells, micro-turbines, combustion turbine, and internal combustion engine. [21]

Typical power capability ranges of various DG sources are shown in Table 2.3 which is taken from [21].

| Technology | Typical Capability Ranges | Utility Interface |
|---|---|---|
| Solar, photovoltaic | A few W to several hundred kW | dc to ac converter |
| Wind | A few hundred W to a few MW | asynchronous generator |
| Geothermal | A few hundred kW to a few MW | synchronous generator |
| Ocean | A few hundred kW to a few MW | four-quadr. synchronous machine |
| ICE | A few hundred kW to tens of MW | synchr. generator or ac to ac converter |
| Combined cycle | A few tens of MW to several hundred MW | synchronous generator |
| Combustion turbine | A few MW to hundreds of MW | synchronous generator |
| Microturbines | A few tens of kW to a few MW | ac to ac converter |
| Fuel cells | A few tens of kW to a few tens of MW | dc to ac converter |

**Table 2.3 Typical power capability ranges for various DG technologies** [21]

As evident from Table 2.3, combustion turbines have the highest power producing capability, whereas solar and combined cycle turbine technologies have a relatively less capability to produce power.

### 2.2.4 Applications of Distributed Generation

DGs have a wide range of applications. They can be used as a standby source of power in case the main grid fails. It can then supply power to sensitive loads in a normal way. This is commonly called stand-by operation mode. Isolated areas can use distributed grids as it might not be feasible to connect the grid in those geographically challenging areas.

### 2.2.5 Distributed Generation Challenges

The types of challenges [22] which are being faced by DG can be divided into two major classes:

**Technical Challenges**: They pertain to the issues like safe operation, emergency operation (island mode), power quality and stability issues.

**Non-technical Challenges**: They relate to the factors like cost of purchase, insurance and setting standards for interconnection.

### 2.2.6 Major Policy Issues

There are some major policy issues regarding distributed generation. Firstly, they are expensive to install and maintain. Their protection schemes are relatively difficult to design. In the event of islanding, special considerations must be taken into account such as no power is being supplied to the grid at the instance of outage. Secondly, some DG units use induction generators which are not capable of producing reactive power in the system. We know that power usually flows unidirectional from high voltage level to a low one i.e. from transmission grid to distribution grid. Increased number of DG units may result in power flows from the low voltage to medium voltage grid. Some DG technologies produce direct current, so, these units must be connected to the grid using inverters and DC-AC interfaces. This can give rise to harmonics which disturbs the quality of output power. It is essential to filter these harmonic distortions properly otherwise they may affect the operational capabilities of the load. [18]

## 2.3 Microgrids

According to the U.S. Department of Energy, microgrid is "a group of interconnected loads and distributed energy resources within clearly defined electrical boundaries that acts as a single controllable entity with respect to the grid (and can) connect and disconnect from the grid to enable it to operate in both grid-connected or island-mode." The basic function of micro-grid is to maintain stable operation under various faults and factors which can disturb the network stability. In simple words, it provides power to the local load. In other words, micro-grid is a smaller version of main grid which operates independently. [23], [24]

### 2.3.1 Modes of Operation

Micro-grid has two modes of operation which are highlighted below. [25]

**Grid Connected Mode**

Under this mode, main grid is active. Static switch is closed. All feeders are being supplied by main grid. In other words, critical loads (on Feeders A, B & C) and non-critical loads (on feeder D) are being supplied by the main grid. Figure 2.4, taken from [25], displays this scenario.

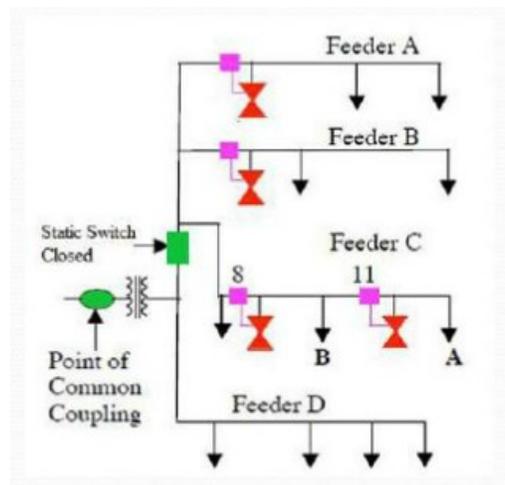

**Figure 2.4 Grid-Connected Mode** [25]

**Grid Disconnected Mode**

It is also called island mode. Under this condition, main grid is cut off and is not supplying power. Static switch is open. Feeders A, B and C are supplied by micro-sources. Feeder D is dead as it is not sensitive. In other words, critical loads are being supplied power by the microsources and non-critical loads (on Feeder D) are cut-off from the system. Figure 2.5, taken from [25], displays this situation.

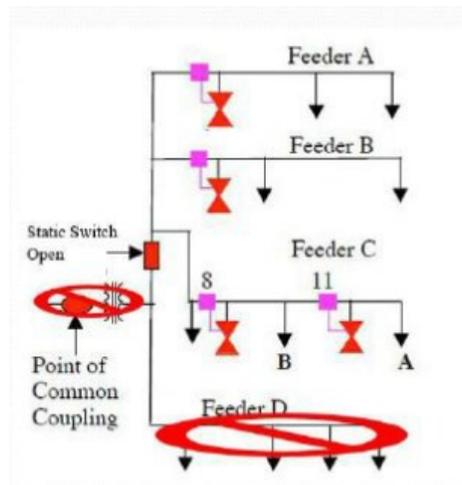

**Figure 2.5 Grid-Disconnected Mode** [25]

Basic micro-grid structure, taken from [12], is shown in Figure 2.6.

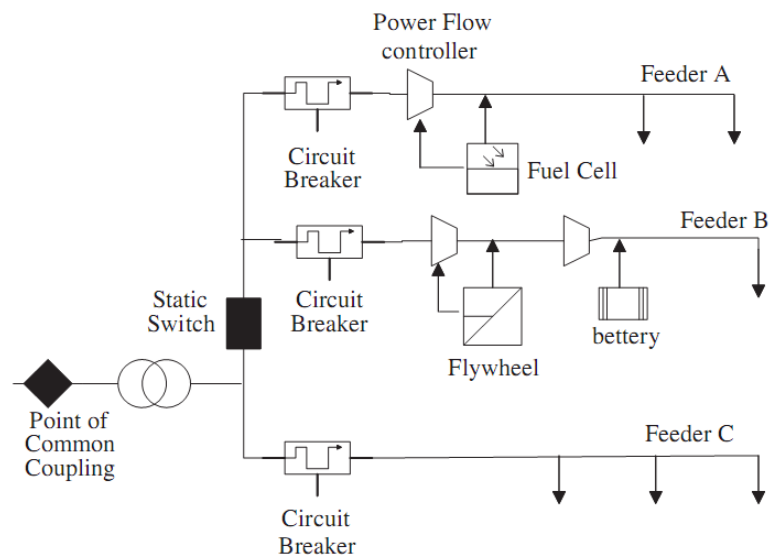

**Figure 2.6 Basic micro-grid structure** [12]

Referring to Figure 2.6, fuel cell is the source of distributed generation and is supplying power when the static switch opens. Circuit breaker is incorporated to trip in the event of fault.

## 2.3.2 Types of Micro-grid

There are many types of micro-grids some of which are explained below. [26], [28]

**Remote grids**: These are the grids which have their existence due to the rugged and severe topographical features of landscape. They are used for locations which are quite far away from the power grid. They are employed when it is not possible to connect all loads to a single main grid. In other words, these grids are used to supply power to geographically challenging areas.

**Military grids**: They are also known as security grids. They are used to keep a record of important data pertaining to security etc.

**Commercial grids**: They are also called industrial grids. They refer to the grids which cater for the energy needs of large industries and factories. They are employed due to their increased security and reliability. They are most commonly used in chemical industry and chip manufacturing industry.

**Community grids**: They also known as the utility grids. They are the most commonly used now-a-days. They refer to the grids which cater for the energy needs of domestic consumers.

Figures 2.7 & 2.8, taken from [26], show pictorial representations of commercial (industrial) grid and community (utility) grid.

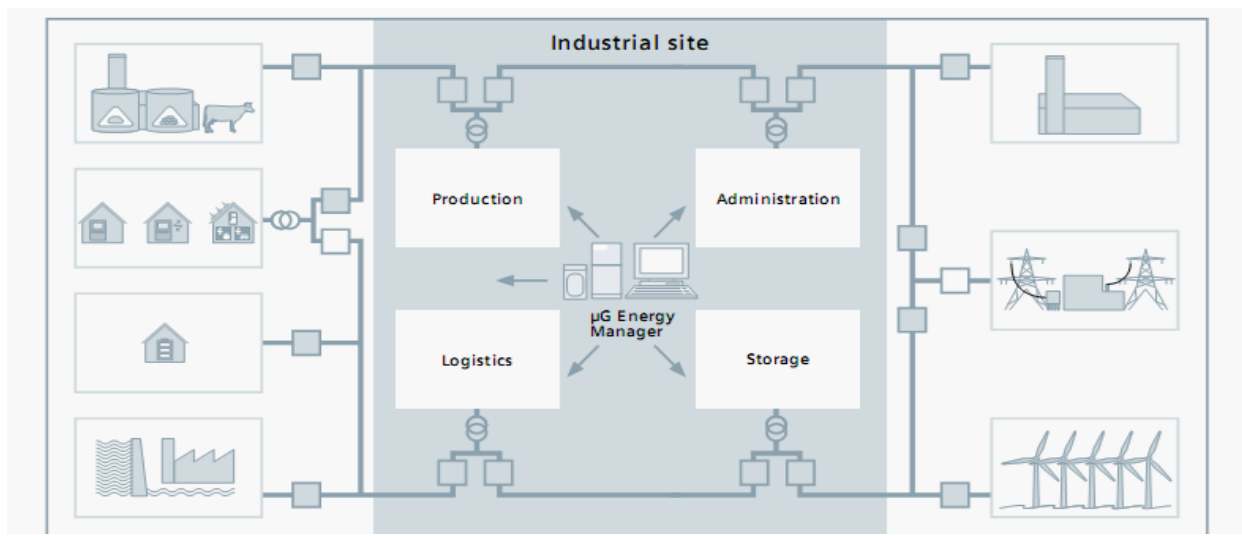

**Figure 2.7 Industrial Microgrid** [26]

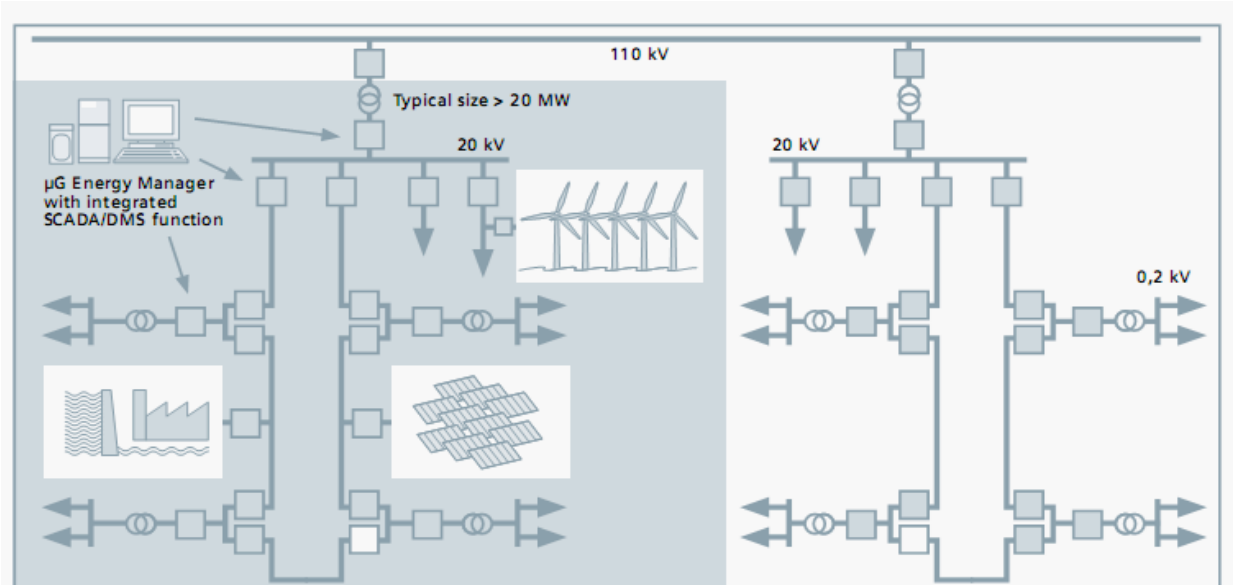

**Figure 2.8 Utility Microgrid** [26]

### 2.3.3 Features of Microgrid

Some major features of micro-grid according to literature [15], [16]:

(1) Peer-to-peer environment (Decentralized control).

(2) Plug-and-play (Utilization of waste heat from gases to improve efficiency).

(3) Smooth seamless transfer between island mode and grid-connected mode.

### 2.3.4 Stages of Operation

Micro-grids have four stages of operation [6]:

(1) Transient stage of going to grid-connected mode.

(2) Steady stage of grid-connected mode.

(3) Transient stage of going to island mode (grid-disconnected mode).

(4) Steady stage of island mode (grid-disconnected mode).

## 2.3.5 Differences between Microgrid & Main grid

Literature [6] has pointed out some major differences between traditional main grid and microgrid. These are:

(1) To the main network, microgrid is a modular controllable unit, just like a generator or load, which can be scheduled. But to the consumers, it is an autonomous system, which can satisfy different power quality requirements from different loads. It basically acts as a source of power to local loads.

(2) There are many unconventional generators in a microgrid like wind power, photovoltaic or fuel cells.

(3) Besides power, microgrid has the ability to supply heat. It can utilize waste heat using CHPs, hence, giving a rise to the overall efficiency of the network.

## 2.3.6 Advantages of Microgrids

Some advantages of using micro-grids are [17], [24], [26], [27]:

(1) They lessen the price of the electricity by providing their power to the main grid. This results in sufficient reduction of required power demand.

(2) Combined heat and power distributed generation can utilize the waste heat from gases for increasing the overall efficiency of the system.

(3) They can easily be manufactured in the form of modules and hence, can be installed quickly at any suitable location.

(4) As there are many DG technologies, there is no need of using a particular kind of fuel or DG source. Hence, there is a large variety and choice in the use of micro-sources.

(5) They increase system reliability in case of a fault on the main grid. They ensure continuity of supply to the consumer end. It is practically made possible by the rapid transfer between grid-connected and island mode.

(6) They employ renewable sources like wind and PV as micro-sources. Hence, they don't pose serious threats to environment. They are environmentally friendly.

(7) In the long term, this area is a potential job creator for micro-grid construction, installation and maintenance.

(8) As the micro-grid involves placing of power sources close to the load, there are minimum transmission and distribution losses.

## 2.3.7 Key Issues of Microgrids

There are some key issues regarding micro-grids which must be addressed. [24], [25]:

- ✓ It is very vital in a power system to balance the real and reactive power. Such imbalance happens when power generated is not equal to the power demand. This can cause the system frequency to drift from its nominal value. This imbalance can be dealt with the help of frequency and voltage control. That is why, active power-frequency droops and reactive power-voltage droops are used.
- ✓ Another issue regarding the stability of micro-grids is their operation under island mode. The micro-grid must provide power to all the loads which are controlled by the micro-sources.
- ✓ Resynchronization with the main grid. This is difficult and requires a lot of care. It is because the voltages and frequency, before closing the switch, must be equal to the steady state values of voltages and frequency.

# 2.4 Recent Development Projects regarding Microgrids

This section gives a brief description about some of the recent microgrid development projects in some countries:

## 2.4.1 Japan

In 2003, the new Energy and Industrial Technology Development Organization and the Ministry Of Economy, Trade and Industry, Japan started three practical projects which dealt with introducing renewable energy resources into local power grid. The main target which was achieved by them was the design of a system which had efficient control and reliability. As this was not economically feasible, thus, some methods to improve the economics of the designed system were presented. [24]

### 2.4.2 Canada

In Canada, Micro-grid R&D has done some work on medium voltage level. Here, they have placed their stress on designing and developing new protection and control techniques and investigated them in detail, especially, when there is a large input of wind power into the system. They have looked into both modes of operation i.e. grid-connected mode and island mode. [35]

### 2.4.3 India

Similarly, in India, a hilly area by the name of Alamprahu Pathar (in the state of Maharashtra), was selected as a potential site for carrying out a micro-grid project. This site has good generation of wind power and construction of a micro-grid here can benefit nearby sugar industries as well as the domestic and agricultural consumers in the nearby area. [35]

### 2.4.4 China

In Xiamen, China, a microgrid working on direct current will start functioning by the end of this year. The Haixi International New Energy Industry Expo, Nextek Power Systems, People Power Company, Canadian Solar, Intel Corporation and Lawrence Berkeley National Laboratory have announced that they will manage, design and develop this microgrid at the School of Energy Research in Xiamen University, in Xiamen, China. This plan will further boost the economy of China and provide a good support to the energy sector. It will result in significant energy savings and will enhance the operational efficiency of the electrical power networks. [36]

### 2.4.5 Lithuania

In recent years, some trends in moving towards distributed generation have been seen in Lithuania. In the last decade, the country has developed many combined heat and power plants to tackle the microgrid issue. However, there are still some major hurdles before the system of microgrid generation can be fully implemented in this country. Appropriate legislation needs to be done regarding this. There are also some technical barriers which prevent this for example small generators need to meet some criteria which are set by the network operator. There are also some issues related to connection validation time. If these issues are resolved, this country can progress rapidly in the energy sector. [37]

# CHAPTER 3

# SIMULATIONS INVOLVING MAIN GRID AND SYNCHRONOUS GENERATORS

This chapter presents simulation work involving grid, synchronous generators and various loads. Behaviour of various electrical parameters is observed. Some insight into load transients is also provided.

## 3.1 SimPower Systems

The software which is used to carry out the simulation work is called SimPower Systems. SimPower Systems is basically a part of MATLAB library. It is a very useful tool for simulation, modeling and evaluation of electrical power networks. The SimPower library consists of a large range of electrical power circuit elements, for example, three-phase sources, transmission lines, circuit breakers, fuses, transformers, switches, diodes and a variety of loads. The salient features of this software also include wind generators, AC drives and Flexible AC Transmission System (FACTS). This software operates in conjunction with Simulink software to model and evaluate electrical power systems. In other words, SimPower operates in the same environment as Simulink. [29]

### 3.1.1 Choosing the Solver

Simulink has a variety of solvers in order to solve an electrical network. The solver basically determines the step time of the simulation and the way the simulation proceeds. This choice depends on various factors.

ODE45 is the most commonly used solver. It uses Runge-Kutta and Dormand-Price integration techniques for evaluating step times. Its accuracy is also quite reasonable. In case, the problem or the network is stiff, it is recommended not to use ODE45. Stiff network means the network whose differential equations cannot be solved by a particular method and is numerically unstable. If such is the case, ODE23 is preferred. It is relatively accurate and faster when the circuit is stiff. There are some other solvers like ODE15s and ODE113 but they are not too common and are used in some special cases and problems which are computationally exhaustive. [30]

## 3.2 Simulation of network involving main grid and a synchronous generator

Firstly, the system consisting of a main grid, a synchronous generator and resistive load is investigated. The block diagram is shown in Figure 3.1. The actual Simulink diagram and its subsystems are shown in Appendix A1.

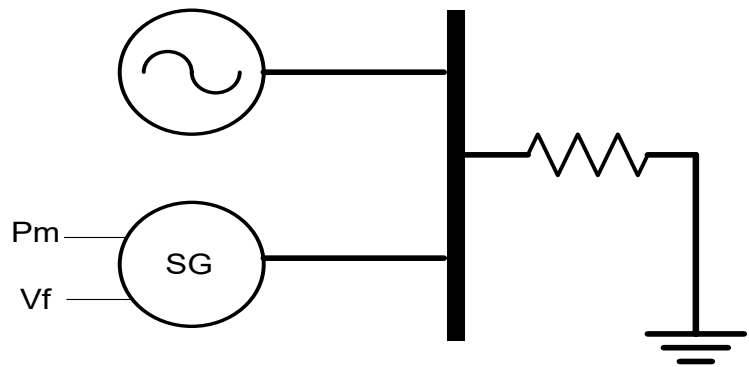

**Figure 3.1 Block diagram showing the connection of grid, synchronous generator and load**

Refer to Figure 3.1. The three-phase voltage source acts as the grid. Its rated line to line rms voltage is 11 KV. It is rated at a frequency of 50 Hz. Its internal resistance and inductance is 0.00001 Ohms and 0.04 H respectively. It is internally grounded.

The synchronous generator is rated at 1.5 MVA, 11 KV (line to line rms) and 50 Hz. It has 4 poles and has a cylindrical rotor. It has 2 inputs. One of them can either be mechanical power $P_m$ or the speed and the other one is the field voltage $V_f$. In Figure 3.1, mechanical power $P_m$ and field voltage $V_f$ are the inputs. The input mechanical power is 0.5 MW. Field voltage is controlled using a suitable PI controller in such a way that $V_f$ produces nominal terminal voltage (which is 11 KV in this case).

At t=0, generator is connected to the system and it starts to operate as soon as the simulation starts.

The load is resistive. It is adjusted to absorb a load power of 1 MW. The line to line bus rms voltage is nearly 11 KV.

The speed of the synchronous generator is 157 radians/second. It is because the main grid (three-phase voltage source) sets the frequency of the generator in this case and hence, the generator runs at 157 radians/second. In technical terms, the grid is acting as the "Master" and the synchronous generator is acting as the "Slave".

The three-phase measurement block is used for measuring the currents and voltages in the three lines. It is connected in series with a three phase element. It can measure phase-to-ground as well as phase-to-phase voltages and currents. The three phase voltages and currents can be easily converted to the $\alpha\beta$ frame for calculating grid active and reactive powers. The block is shown in Figure 3.2

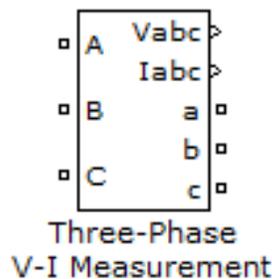

**Figure 3.2 Three Phase Measurement Block**

## 3.2.1 Simulation Results and Discussions

The network in Figure 3.1 is simulated for 15 seconds and the solver used is ode23tb. We know that the formula used to calculate the speed of synchronous generator is given by:

$N_s = 120\frac{f}{P}$

Where, $P$ is the number of poles and $f$ is the system frequency. In the given system, $f$ is 50 Hz and number of poles are 4, so, the synchronous speed is 1500 RPM (157 radians/second).The synchronous generator takes in 0.5 MW of active power. In other words, we are forcing it to produce 0.5 MW at the output. The load power is already adjusted to absorb 1 MW, thus, the grid must produce 0.5 MW to keep the powers balanced. The results are as expected. The active and reactive powers of the grid (three-phase voltage source), synchronous generator and the load was scoped and observed. It was found that active powers of the grid and synchronous generator are 0.5 MW each which sums up to the load active power (which was preset at 1MW).

## 3.2.2 Graphical Results regarding Active and Reactive Power

The graphical results of active and reactive powers across the generator and the grid are shown in Figures 3.3-3.6

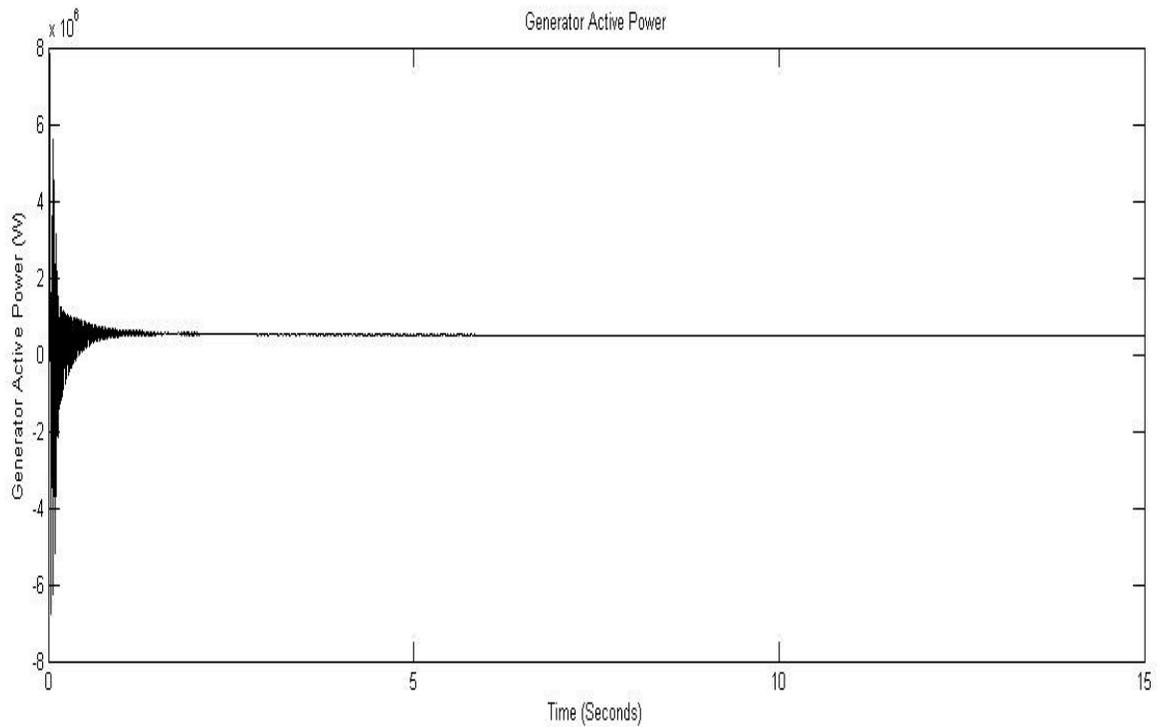

**Figure 3.3 Generator Active Power**

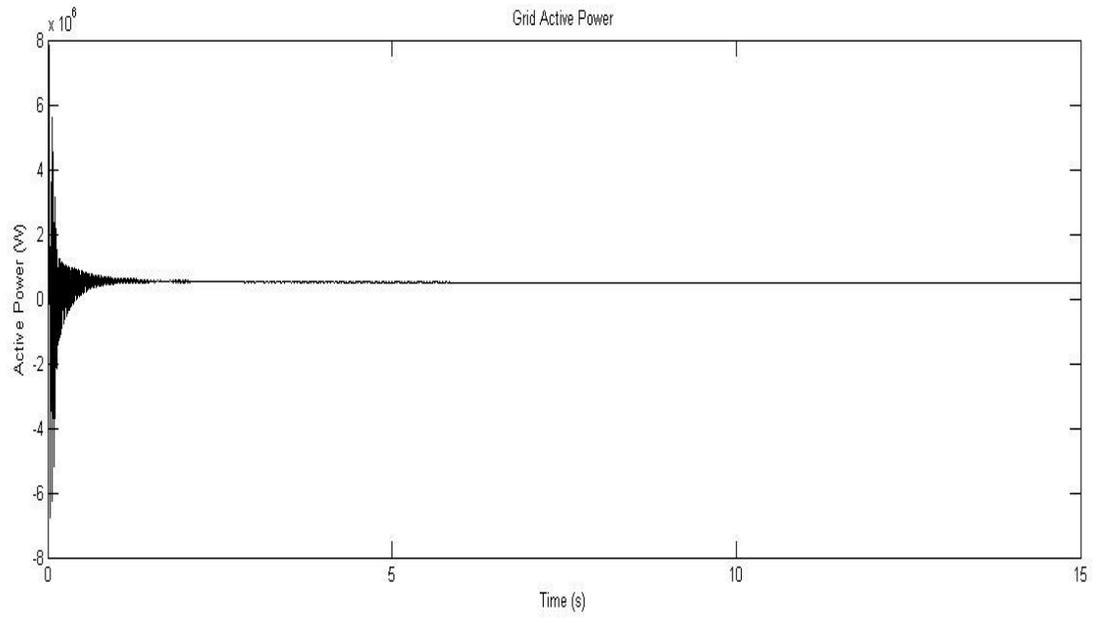

**Figure 3.4 Grid Active Power**

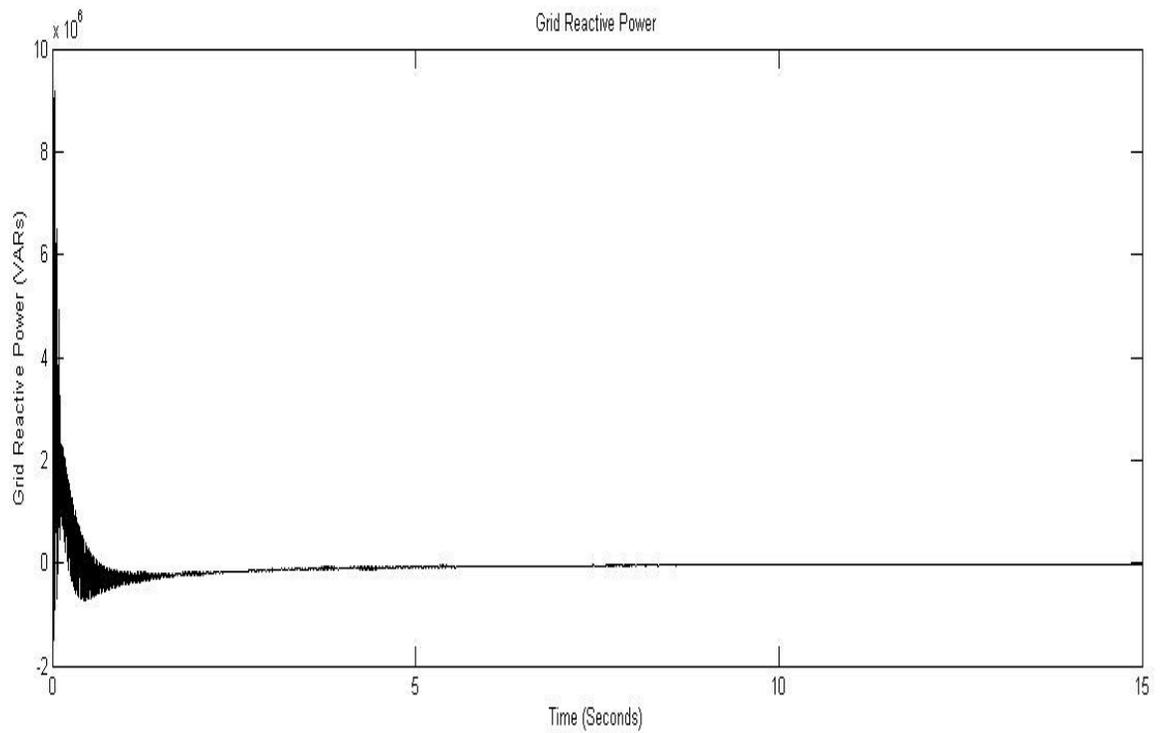

**Figure 3.5 Grid Reactive Power**

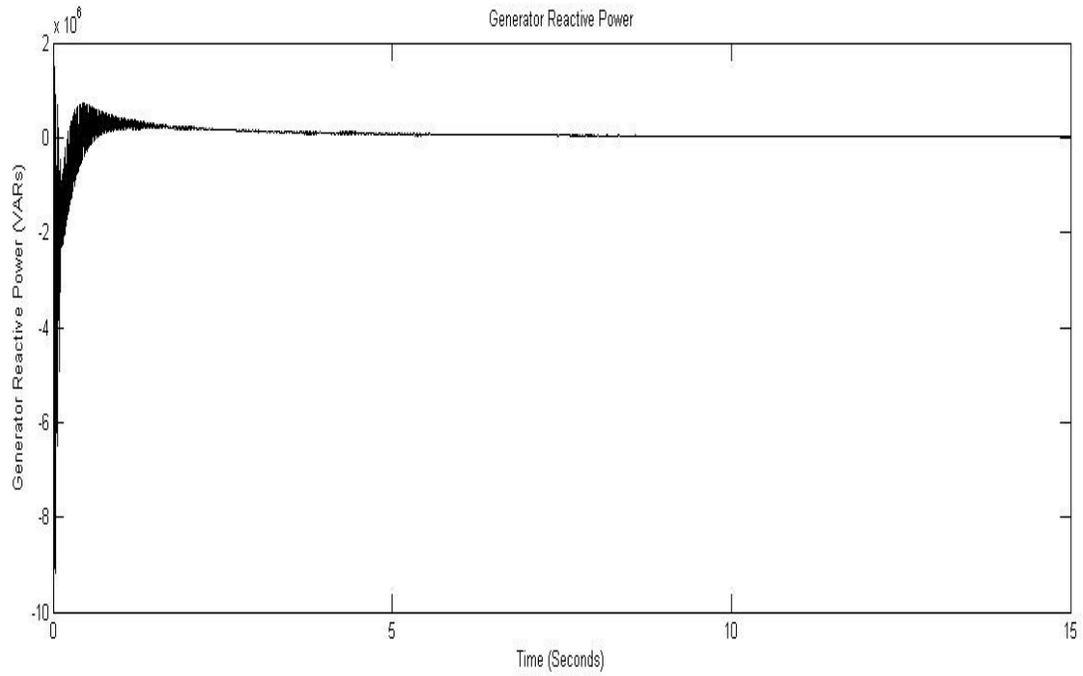

**Figure 3.6 Generator Reactive Power**

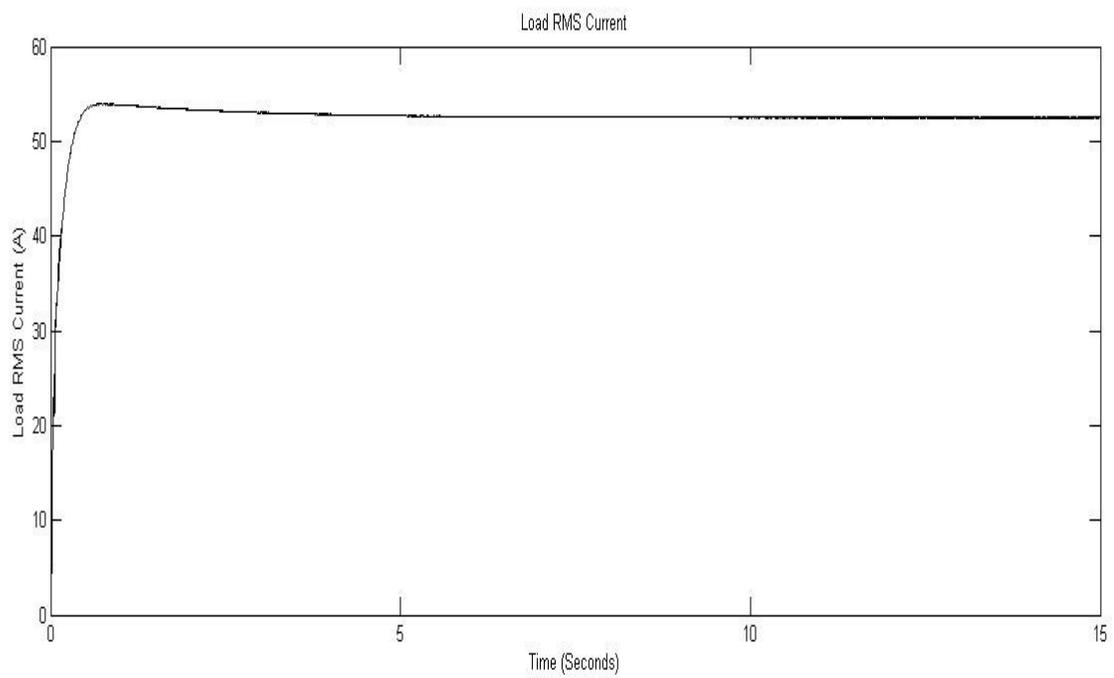

**Figure 3.7 Load RMS Phase Current**

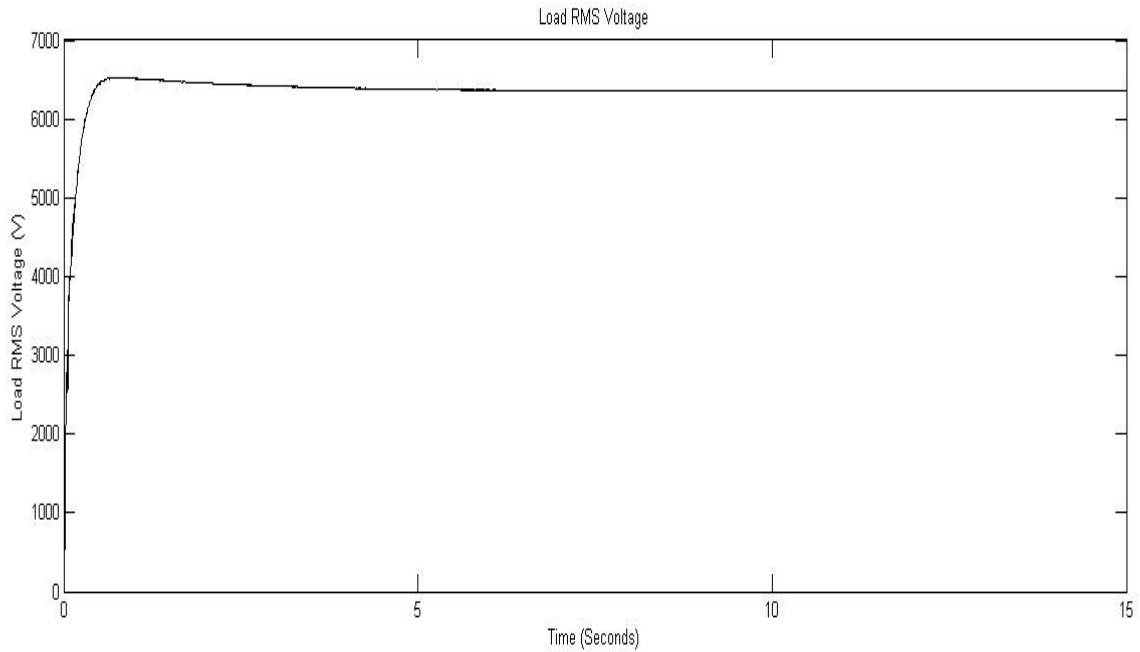

**Figure 3.8 Load RMS Phase Voltage**

### 3.2.3 Discussions regarding Active and Reactive Power

Let us first observe the behavior of active power across the grid, generator and the load. From Figures 3.3 and 3.4, generator and grid active powers are at 0.5 MW each. The load power is at 1 MW which is calculated by the formula:

P= $3 \times V_{rms} \times I_{rms} \times \cos\phi$

Where $V_{rms}$ is the load rms phase voltage, $I_{rms}$ is the load rms phase current and $\cos\phi$ is the power factor (which is unity in this case as the load is resistive).

$V_{rms}$ and $I_{rms}$ from Figures 3.8 and 3.7 are 6353 Volts and 52.5 Amperes respectively.

Hence, load active power P= $3 \times 6353 \times 52.5 \times 1$= 1 MW

Therefore, generator and the grid each supplies 0.5 MW which sums up to the load power of 1 MW.

From Figures 3.5 and 3.6, the grid and generator reactive powers are -36 KVARs and 36 KVARs respectively. The load reactive power is zero as it is a purely resistive load. Hence, reactive power generated by the synchronous generator is being cancelled by that of grid. In other words, synchronous generator is supplying KVARs (reactive power) and the grid is absorbing them.

## 3.2.4 Associated Graphical Results

Some other parameters which are worth looking at are the generator speed, bus voltage and flux along d axis. The generator speed is 157 radians/second and the line to line rms bus voltage is 11 KV. Flux along d axis is about 28 Webers. These are shown graphically in Figures 3.9-3.11.

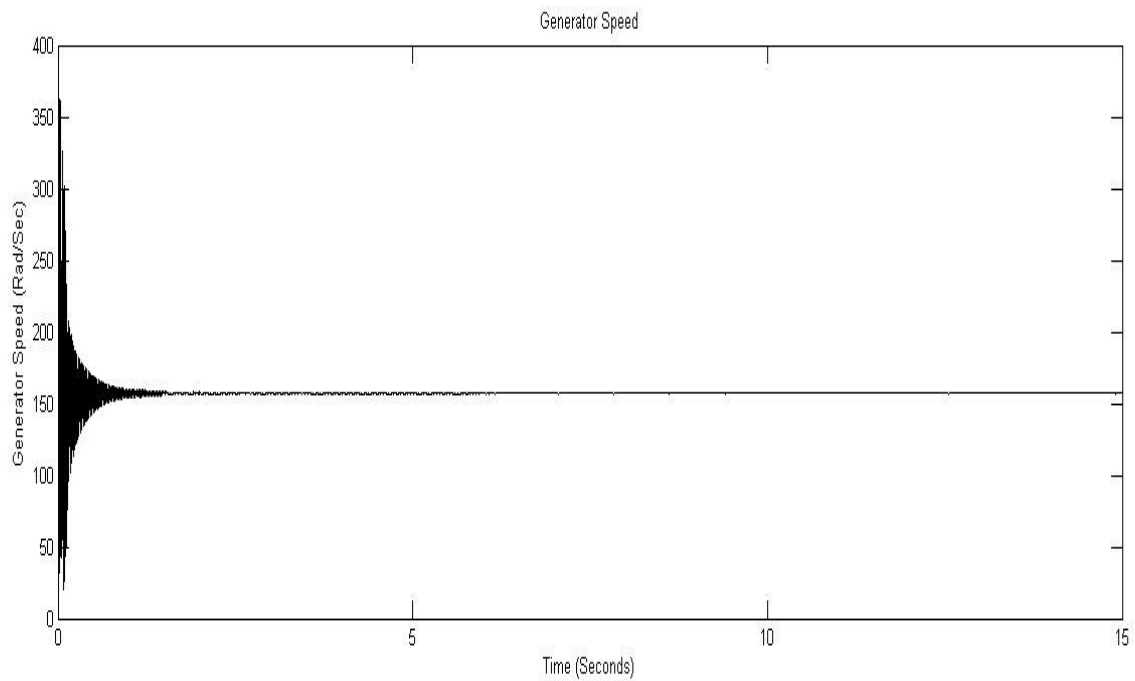

**Figure 3.9 Generator Speed**

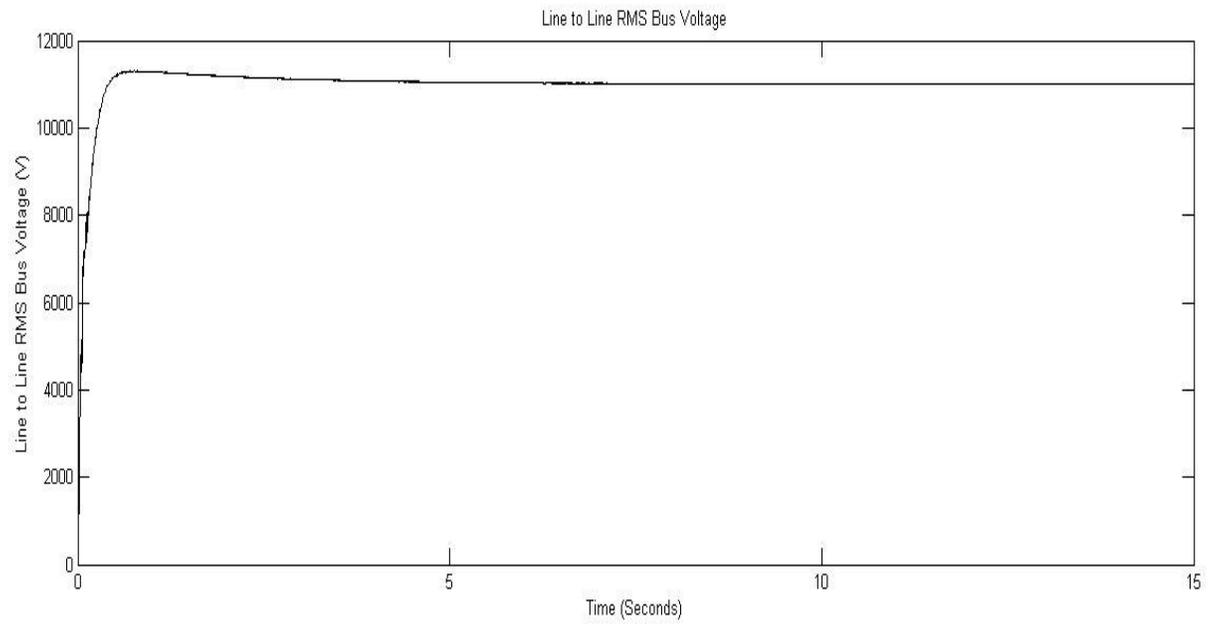

**Figure 3.10 Line to line rms bus voltage**

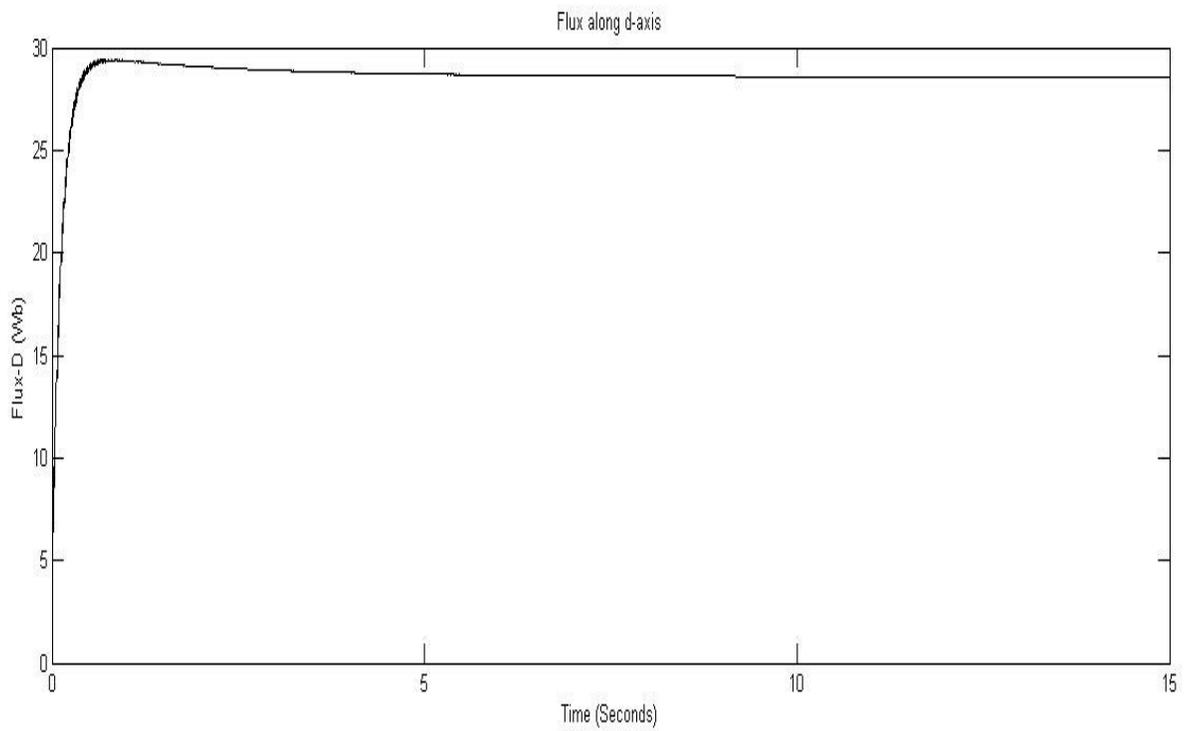

**Figure 3.11 Flux along d axis**

### 3.2.5 Investigation of Load Angle

Load angle is basically the angle between the bus voltage vector (V) and the no load voltage vector E. It is commonly denoted by $\delta$. It takes a very long time to get a steady state value of this load angle for the system shown in Figure 3.1. This problem is discussed in Appendix A2.

## 3.3 Behaviour of load angle for the system involving synchronous generator and resistive load

Now, we shall investigate the trend in load angle in a system which consists of the synchronous generator and resistive load. The block diagram is shown in Figure 3.12. The actual Simulink diagram is shown in Appendix A3.

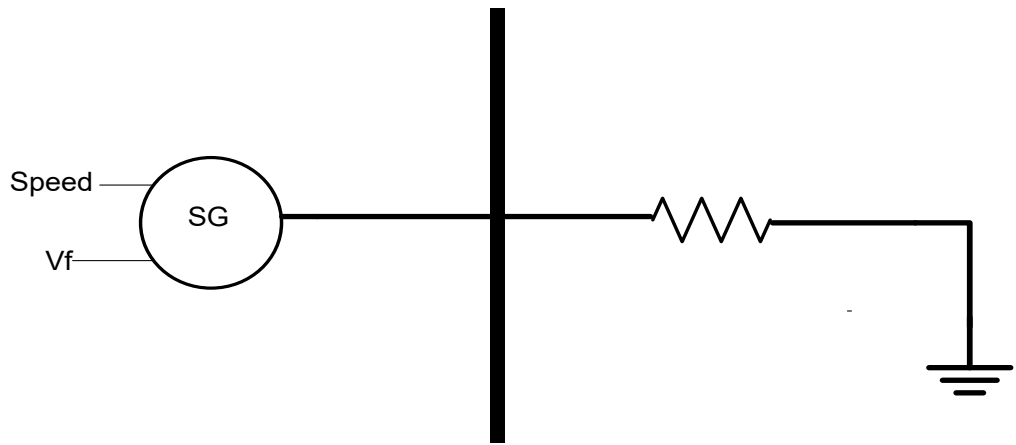

**Figure 3.12 Block Diagram for the synchronous generator and resistive load**

Simulation time used is 1 second and the solver is ode45. All the parameter values for generator and load are exactly the same as for the network in Figure 3.1. The generator is made to run at a constant speed of 157 radians/second. The only difference is that here constant field voltage $V_f$ is applied instead of controlling it through PI controller.

### 3.3.1 Results and Discussions

Values of load resistance were varied and the corresponding load currents and load angles were observed. Results in the tabular form are shown in Table 3.13

| Load resistance (Ohms) | Load current (Amps) | Load angle (Degrees) |
|---|---|---|
| 40 | 20 | 1.16 |
| 50 | 16.2 | 0.96 |
| 70 | 11.6 | 0.73 |
| 15 | 52.96 | 2.815 |
| 5 | 135 | 8.214 |

**Table 3.13 Trends in load angle with variation in load resistance**

As evident from Table 3.13, as the load resistance is increased, load current decreases and load angle also decreases. It is because the smaller the current, the smaller is the $jIX_s$ vector and hence, smaller the angle between the E vector and the bus voltage vector V. Figure 3.14 shows this phenomenon.

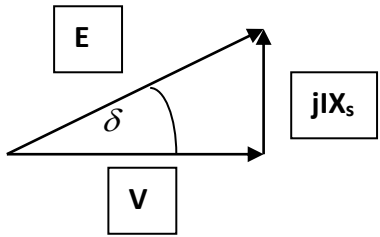

**Figure 3.14 Phasor diagram depicting the relation between load angle $\delta$ and current I**

## 3.4 Behaviour of load angle for the system involving grid, synchronous generator and resistive-inductive load

Now, we shall investigate the trend in load angle for a system which consists of the grid, synchronous generator and resistive-inductive load. The block diagram is shown in Figure 3.15. The actual Simulink diagram is shown in Appendix A4.

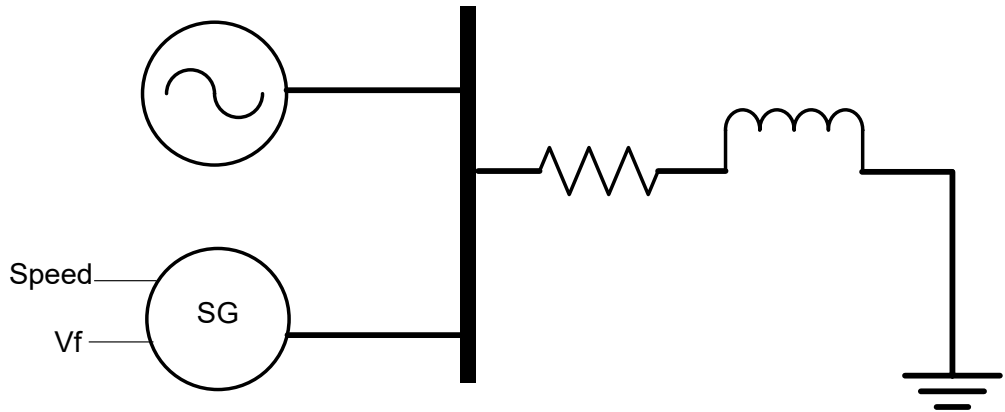

**Figure 3.15 Block Diagram for the grid, synchronous generator and RL load**

Simulation time used is 0.3 second and the solver is ode45. All the parameter values for the generator and the grid (three phase voltage source) are the same as before. Here, inductance of 0.01 H is also introduced along with resistive load. The generator is run at a constant speed of 157 radians/second. Constant field voltage of 70 V is applied.

## 3.4.1 Results and Discussions

The load inductance was kept constant at 0.01 H. Values of load resistance were varied and the corresponding load currents and load angles were observed. Results in the tabular form are shown in Table 3.16

| Load resistance (Ohms) | Load current (Amperes) | Load angle (Degrees) |
|---|---|---|
| 40 | 66 | 1.4 |
| 50 | 57 | 1.2 |
| 70 | 43 | 1.1 |
| 5 | 94 | 4.34 |

**Table 3.16 Trends in load angle with variation in load resistance**

As evident from Table 3.16, as the load resistance is increased, load current decreases and load angle also decreases. It is because the smaller the current, the smaller is the $jIX_s$ vector and hence, smaller the angle between the E vector and the bus voltage vector V.

## 3.5 Investigation of system containing two synchronous generators and resistive/inductive loads

Now, the system consisting of two synchronous generators and resistive/inductive loads will be studied. The block diagrams of the system are shown in Figures 3.17 and 3.18. The actual Simulink diagram is shown in Appendix A5.

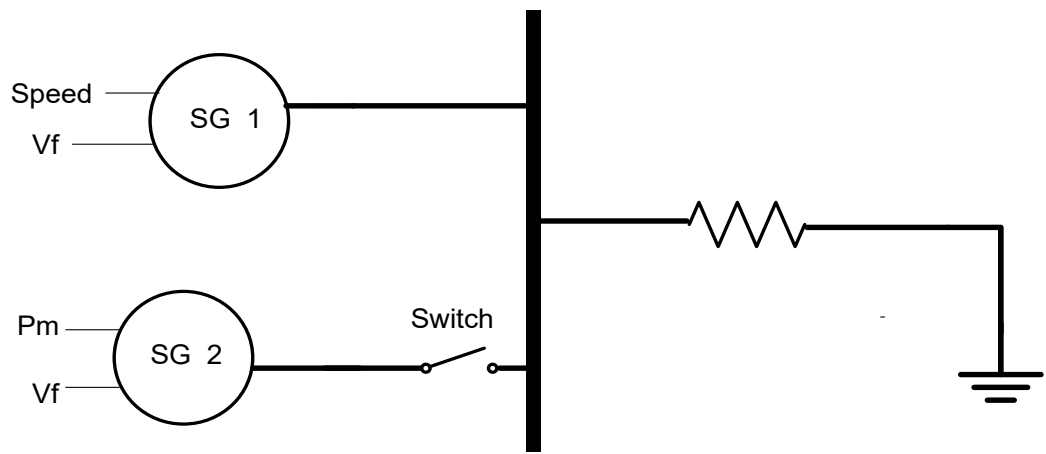

**Figure 3.17 Block Diagram for two synchronous generators and resistive load (Switch open)**

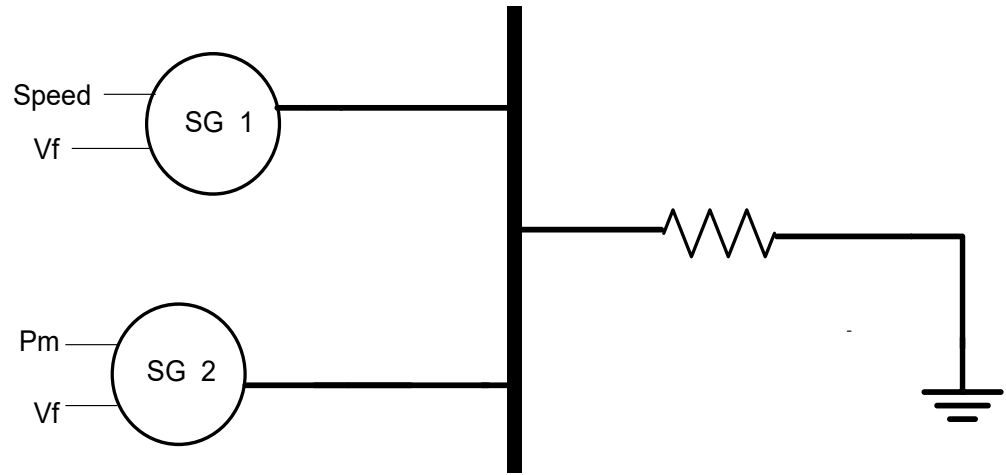

**Figure 3.18 Block Diagram for two synchronous generators and resistive load (Switch closed)**

## 3.5.1 Why Synchronous Generators are operated in parallel?

There are many reasons for operating the synchronous generators in parallel. Firstly, it increases the reliability of the power network. It means that if any generator has to be taken out from the existing system due to some fault, there is not total collapse of load and the other generator can still supply power to the load to some extent. [31]

Secondly, it allows more flexibility in the system. If one generator is to be removed from the system for some maintenance or repair, it can be easily done. Also, this paralleling does not place too much load on the individual generator operation. Last, but not the least, various generators can provide a much bigger load than one machine itself. [31]

## 3.5.2 Conditions for Paralleling

There is some criterion which needs to be satisfied before any generators are to be connected in parallel. They must have same rms line voltages, same phase sequences and same phase angles. The frequency of the generators to be added (usually called the oncoming generators) must be slightly higher than the frequency of the running system. [31]

### 3.5.3 Simulations and Results

For the systems in Figures 3.17-3.18, the simulation time used is 5 seconds and the solver is ode23tb. Both the generators are identical; each rated at 1.5 MVA and 11 KV (line to line rms). The load is resistive and is chosen in such a way that it receives an active power of 1 MW. In this case, SG 1 is acting as the "Master" and SG 2 as the "Slave". In other words, SG 1 sets the speed of SG 2.

Firstly, the switch is off (Figure 3.17) and SG 2 (Synchronous Generator 2) is run at zero power and SG 1 (Synchronous Generator 1) is made to run at a synchronous speed of 157 radians/second. Here, the load power (1 MW) equals the power produced by SG 1 (1 MW). After that, SG 2 is run at 0.5 MW and power sharing is observed. In this case, SG 1 also produces 0.5 MW to keep the active power across load at 1 MW. We can produce any amount of power (within limits) from SG 2 to see the power sharing, for example, if SG 2 is forced to run at 0.7 MW, SG 1 will produce 0.3 MW and so on.

The reactive powers produced by SG 1 and SG 2 are -21 KVARs and 21 KVARs respectively. The net reactive power is zero as the load is purely resistive.

The graphical results for active and reactive powers of both synchronous generators are shown in Figures 3.19-3.22.

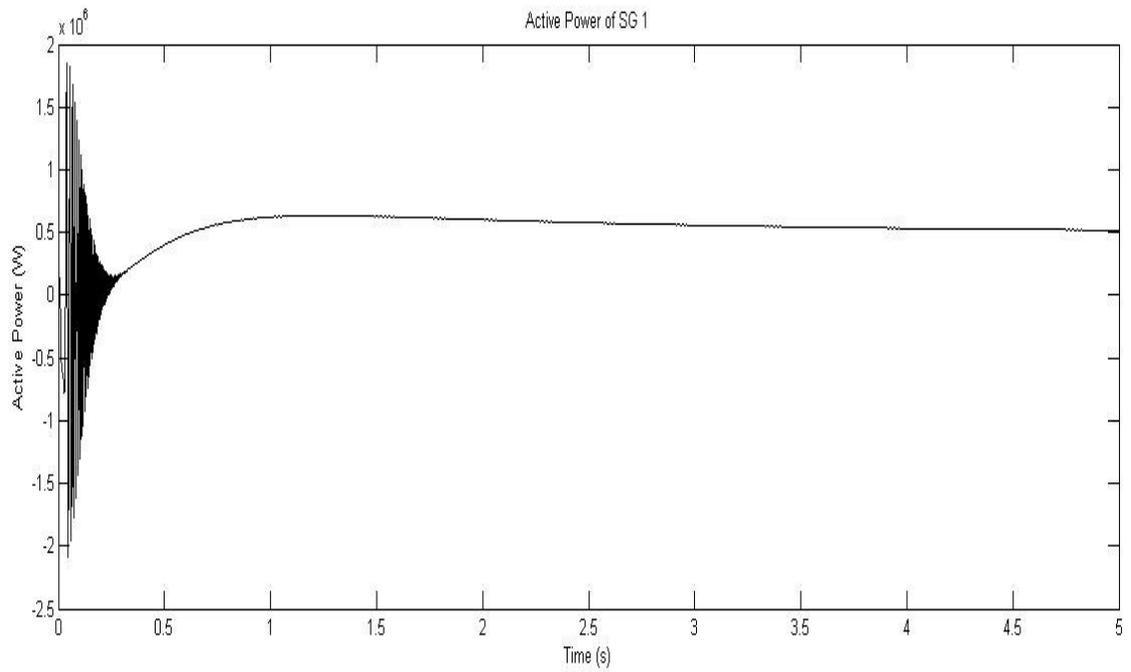

**Figure 3.19 Active Power of SG 1**

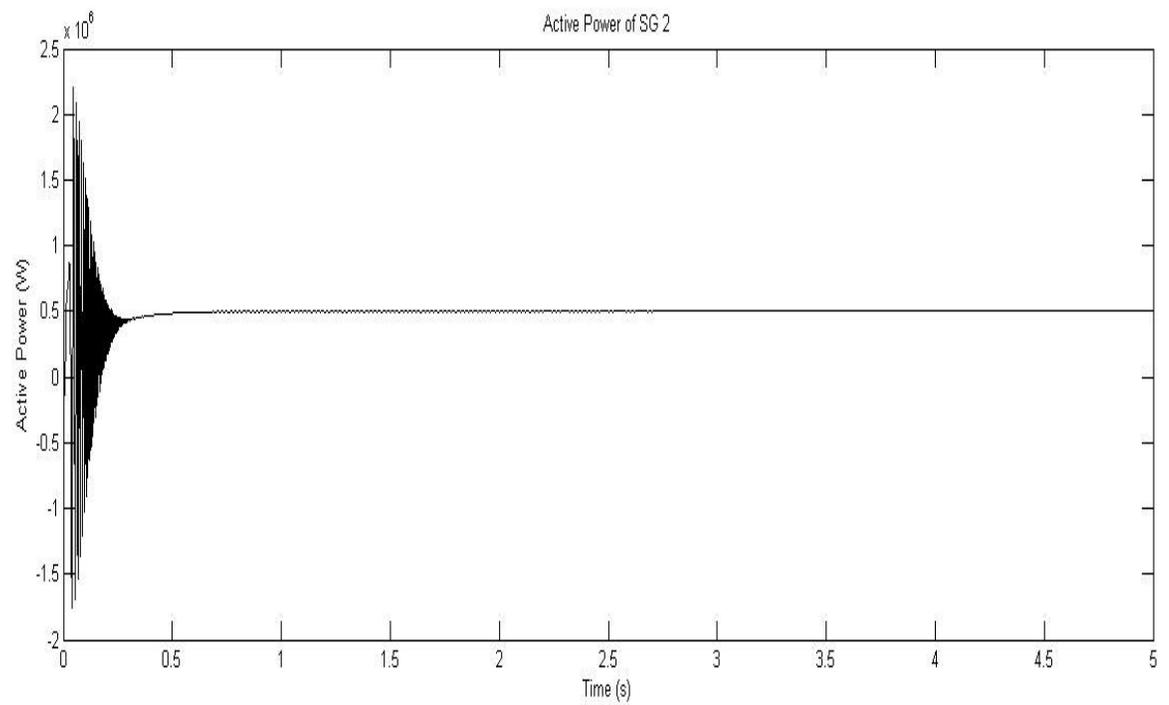

**Figure 3.20 Active Power of SG 2**

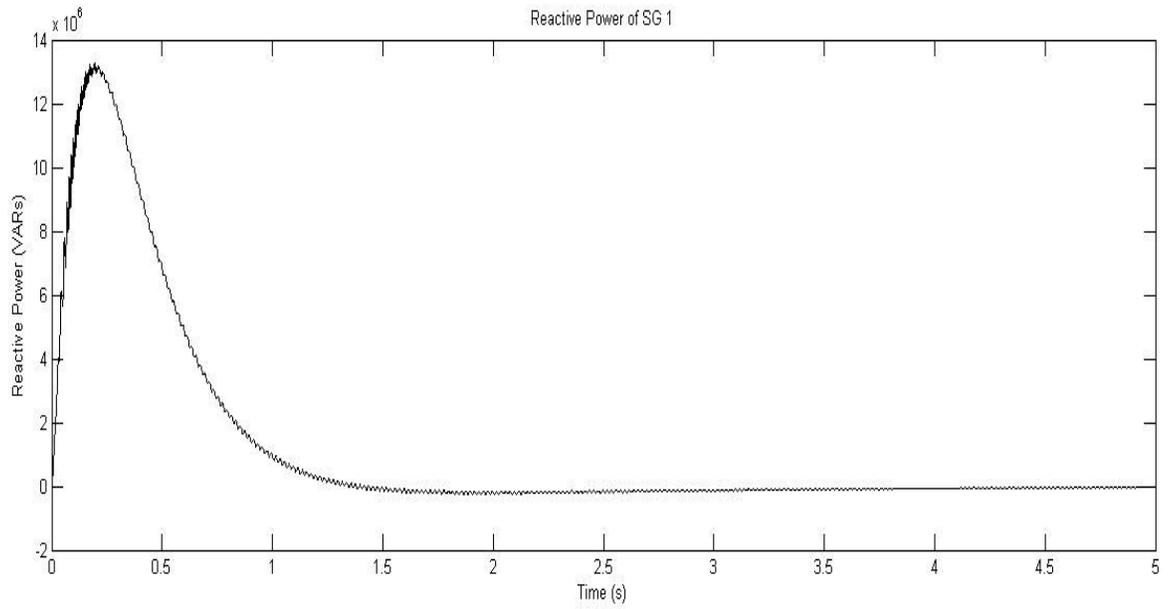

**Figure 3.21 Reactive Power of SG 1**

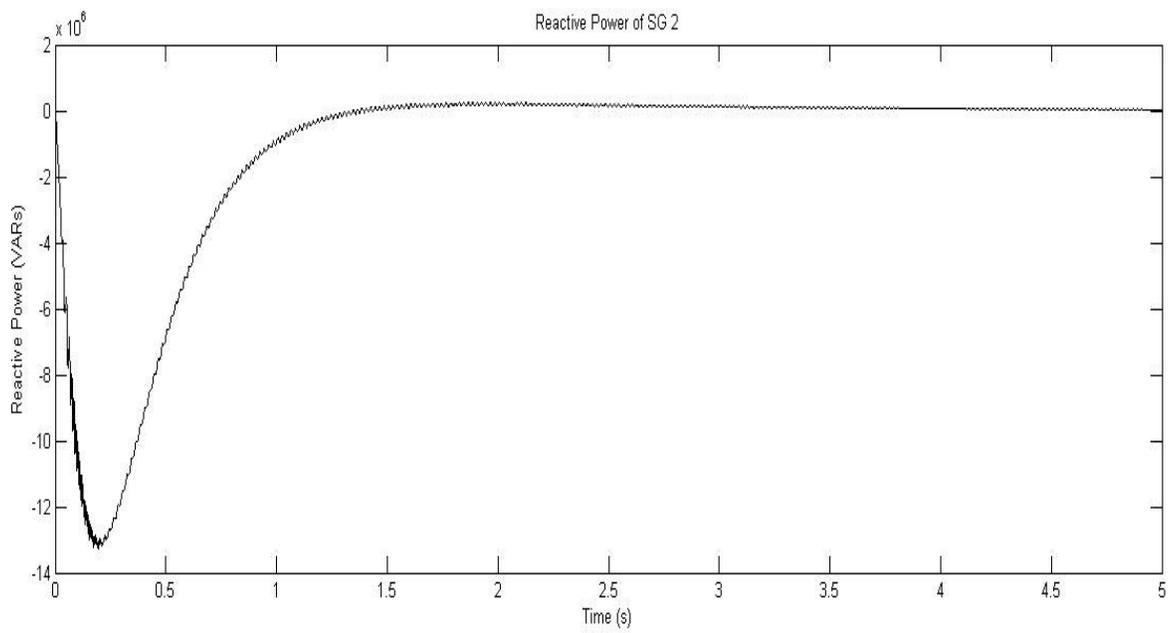

**Figure 3.22 Reactive Power of SG 2**

### 3.5.4 Load Transients

Load transients and their characteristics were also observed for the system of Figure 3.18, for example, a resistive or inductive load comes into the system after some preset time.

Consider the network shown in Figure 3.23. The generator parameter values are the same as in the network of Figure 3.18. The value of load resistance is 121 ohms. The network is simulated for 5 seconds. The solver is ode23tb. The switch is turned on after 2 seconds. In other words, a resistive load comes into the system after 2 seconds. The graphical presentation for this load transient is shown in Figure 3.24.

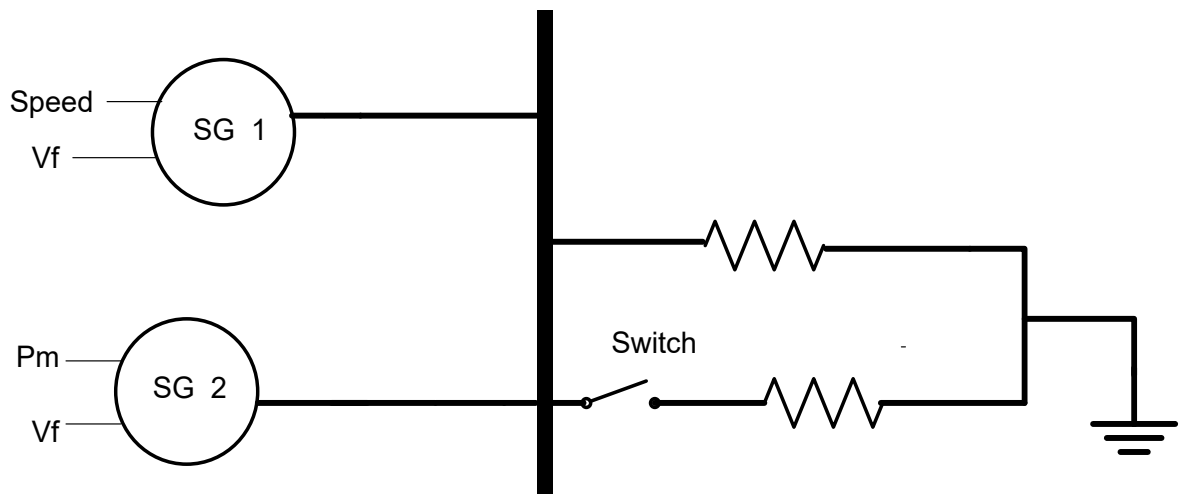

**Figure 3.23 Resistive Load Transient**

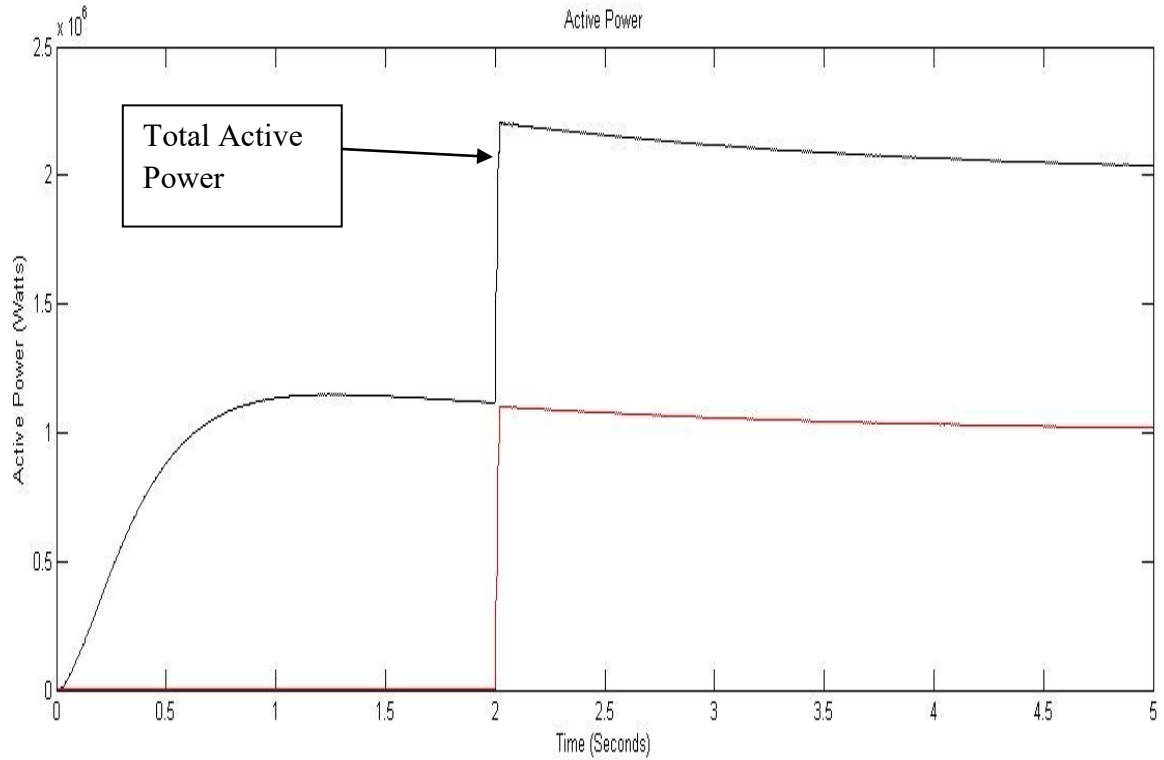

**Figure 3.24 Graphical representation of Resistive Load Transient**

Figure 3.24 shows the graph when resistive load (of 1 MW) comes into the system after 2 seconds. In this case, the total load power is obviously 2 MW. So, SG 1 supplies 1.5 MW and SG 2 supplies 0.5 MW. It is because we are forcing SG 2 to run at 0.5 MW. If, for example, we make it run at 0.7 MW, the other generator will supply the remaining 1.3 MW to make the load active power equal to 2 MW.

Now, consider the network shown in Figure 3.25. Values of load resistance and inductance are 121 ohms and 0.1 H respectively. The network is simulated for 5 seconds. The solver is ode23tb. The switch is turned on after 0.5 second. In other words, a resistive-inductive load comes into the system after 0.5 second. The graphical presentation for this load transient is shown in Figure 3.26.

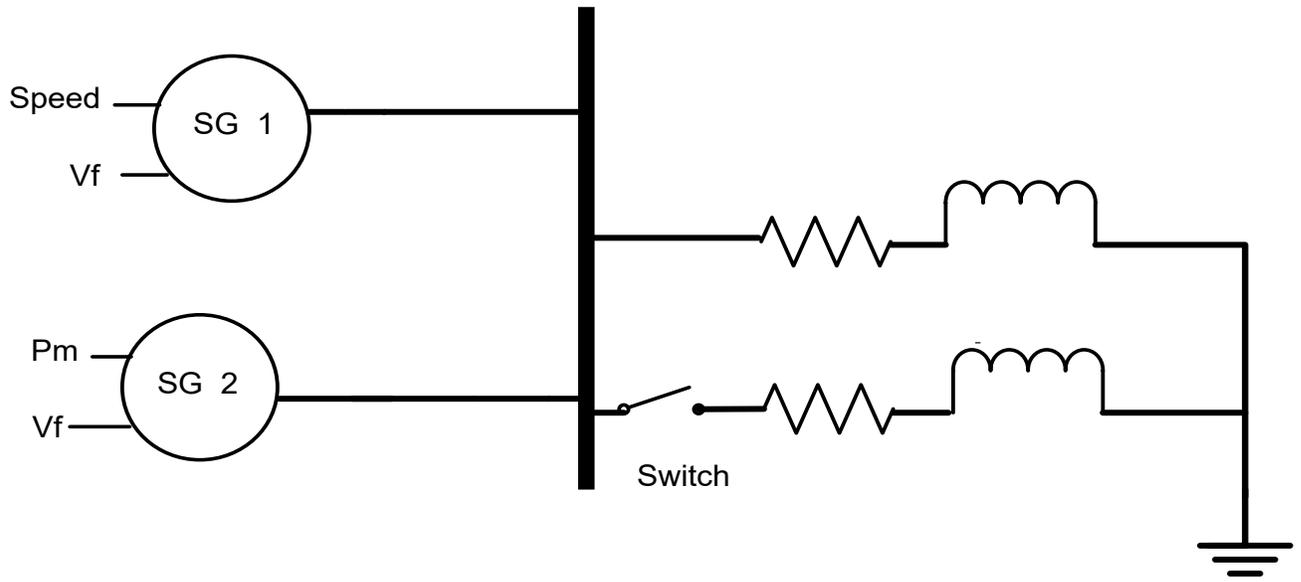

**Figure 3.25 Resistive-Inductive Load Transient**

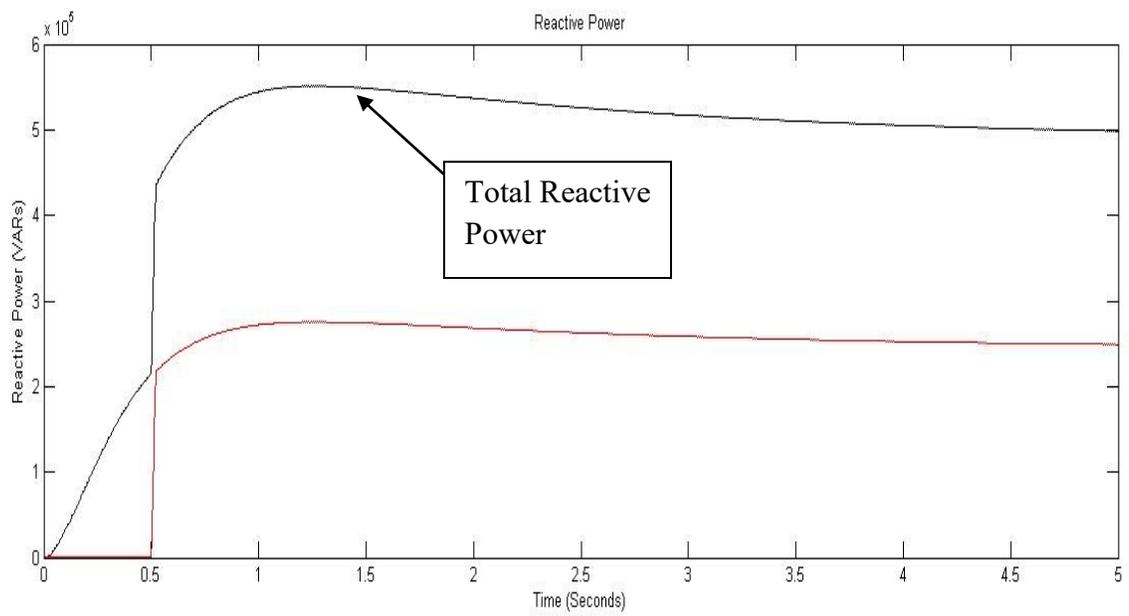

**Figure 3.26 Graphical representation of Resistive-Inductive Load Transient**

As evident from Figure 3.26, inductive load (of 250 KVARs) is added into the system after 0.5 second, thus, the total reactive power is doubled i.e. 500 KVARs.

# CHAPTER 4

# DROOP CONTROLLED SYNCHRONOUS GENERATORS

This chapter discusses the droop control of multiple synchronous generators and how active and reactive power is shared between them. Before that, some theoretical background is discussed.

## 4.1 Droop Control Background

The concept of droop control is very vital when there is more than one synchronous generator connected to the system. Consider the graphs, taken from [32], shown in Figure 4.1

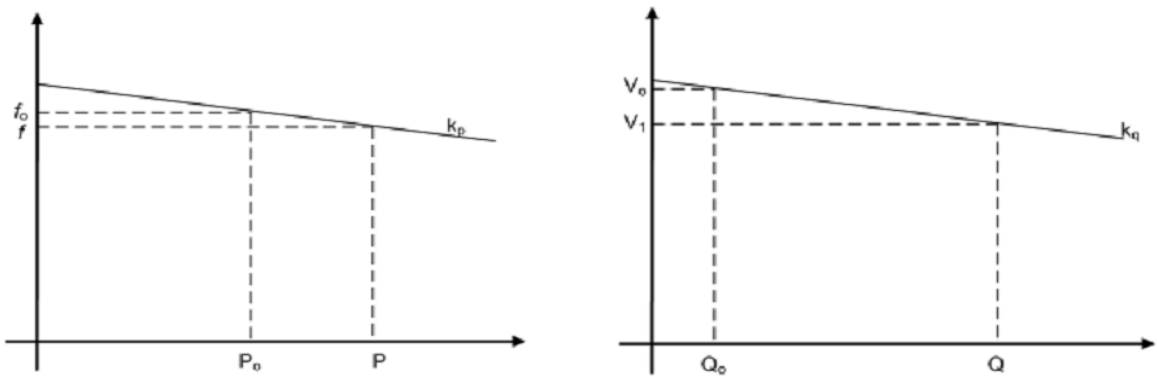

**Figure 4.1 Typical F-P and V-Q droop characteristics** [32]

Figure 4.1 shows the typical active power-frequency and reactive power-voltage droop characteristics. They are commonly known as F-P and Q-V droops. $K_p$ and $K_q$ are the droop coefficients (or gains) of the respective graphs. The higher the coefficient, the smaller will be the active or reactive power, depending on which droops are under consideration. [32]

The well-known equations used to describe these graphs are:

**$f=f_0 -mP$……………………… (4.1)**

**$V=V_0-nQ$……………………… (4.2)**

In these equations, $f_0$ and $V_0$ are the nominal frequency and output voltage magnitude. In other words, they are the values of frequency and voltage at no load. Symbols 'm' and 'n' are the frequency and voltage droops coefficients respectively. [33]

Normally, the difference $(f_0-f)$ and $(V_0-V)$ are allowed to be 2% and 5% range of the nominal values. It must be noted that equations (4.1) and (4.2) hold for mainly inductive impedance, which is usually the case. If the impedance is capacitive or resistive, active power depends on voltage and reactive power depends on frequency i.e. opposite of the inductive case. [33]

All synchronous generators have a source of mechanical power. This source is called the prime mover. The most common type of prime mover is the steam turbine although gas turbines, water turbines and diesel engines are also used as prime movers. Irrespective of what type is the prime mover, they follow a similar trend in such a way that as the power drawn from them increases, the speed at which they operate decreases. This phenomenon is commonly called the speed droop of prime mover. It is defined mathematically by the equation:

**Speed Droop= (X-Y)/(Y)**

Here, X is the no-load prime mover speed and Y is the full-load prime mover speed. [34]

## 4.2 Simulations for Droop Control

The simulations were carried out for the system shown in Figure 4.2. Actual Simulink diagrams and associated control schemes are shown in Appendix B1, B2.

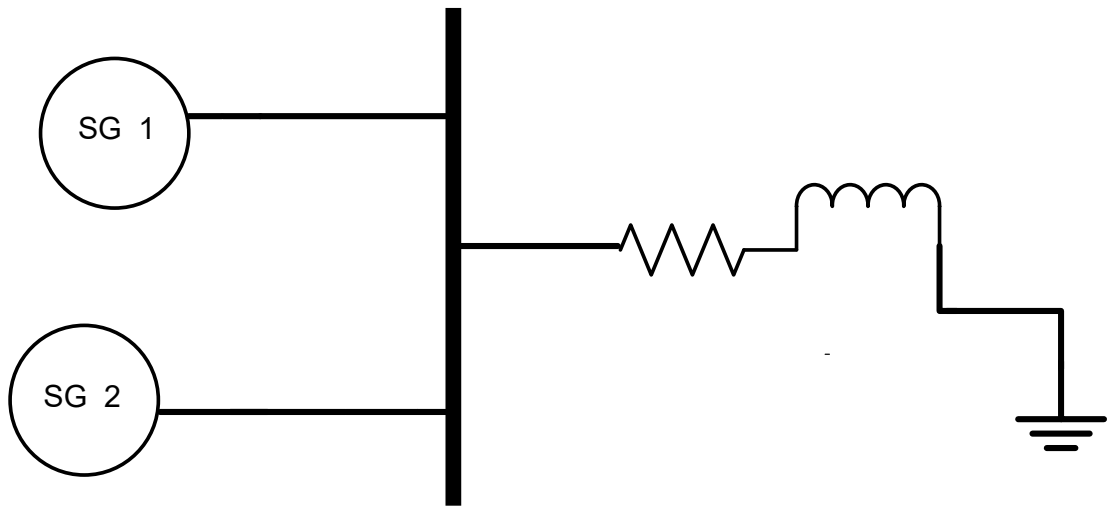

**Figure 4.2 Block diagram consisting of synchronous generators and RL load**

## 4.2.1 Active and Reactive Power Sharing

In SimPower Systems, reactive power–voltage droop and active power-frequency droop control schemes were designed for both generators. In both cases, there is a constant gain in the feedback of the control systems (shown in Appendix B1, B2). These gains are basically the implementation of droop coefficients. These gains determine the active and reactive power sharing between the generators. In other words, the higher these gains (or slopes), the smaller is the active or reactive power contribution by that synchronous generator. The active and reactive powers always equal the load active and reactive power for each case implying that the system is working fine.

## 4.2.2 Results and Discussions

For $V_o$= 7000V, the network in Figure 4.2 was simulated (for 5 seconds with ode23tb as the solver) with appropriate reactive power-voltage droop control scheme (shown in Appendix B1). Some simulation results for investigating reactive power sharing are shown in Table 4.3. K1 and K2 are the droop coefficients (or gains) and Q1 and Q2 are the reactive powers of synchronous generators 1 and 2 respectively. Nominal Q for each generator is 1.05e5 VARs. Load resistance and inductance is 121 ohms and 0.1 H respectively.

| K1 | K2 | Q1 (VARs) | Q2 (VARs) |
|---|---|---|---|
| 0.01 | 0.01 | 1.05e5 | 1.05e5 |
| 0.05 | 0.01 | 3.02e4 | 1.51e5 |
| 0.001 | 0.01 | 2.5e5 | 2.5e4 |

**Table 4.3 Reactive Power sharing for different droop coefficients**

As evident from Table 4.3, the higher the gain K, the smaller the reactive power contribution by that generator and vice versa. It must be noted that the total reactive power (Q1+Q2) equals the load reactive power (210 KVARs) in each case.

For $F_0$= 50Hz (nominal frequency), the network in Figure 4.2 was simulated (for 2 seconds with ode23tb as the solver) with appropriate active power-frequency droop control scheme (shown in Appendix B2). Some simulation results for investigating active power sharing are shown in Table 4.4. K1 and K2 are the droop coefficients (or gains) and P1 and P2 are the active powers of synchronous generators 1 and 2 respectively. Nominal P for each generator is 4.8e5 Watts. Load resistance and inductance is 121 ohms and 0.1 H respectively.

| K1 | K2 | P1 (Watts) | P2 (Watts) |
|---|---|---|---|
| 0.01 | 0.01 | 4.8e5 | 4.8e5 |
| 0.05 | 0.01 | 4.7e5 | 5e5 |
| 0.001 | 0.01 | 4.95e5 | 4.75e5 |

**Table 4.4 Active Power sharing for different droop coefficients**

As evident from Table 4.4, the higher the gain K, the smaller the active power contribution by that generator and vice versa. It must be noted that the total active power (P1+P2) equals the load active power (0.96 MW) in each case.

## 4.3 Variations in Inductive Load

Regarding the reactive power-voltage droop, inductive load was varied and consequently, reactive powers of loads and generators were observed. It was found that higher the load inductance, higher is the reactive power supplied by the generators and their sum equals the load reactive power in each case. Some simulation results are shown in Table 4.5.

| Load Inductance (H) | Q1 (VARs) | Q2 (VARs) | $Q_{load}$ (VARs) |
|---|---|---|---|
| 0.001 | 1555 | 1555 | 3110 |
| 0.1 | 0.1e6 | 0.1e6 | 0.2e6 |
| 0.5 | 0.16e6 | 0.16e6 | 0.32e6 |

**Table 4.5 Reactive power sharing with variation inductance**

As evident from Table 4.5, the higher the value of load inductance, the higher the reactive power supplied by the generators and vice versa. It must be noted that Q1 and Q2 are the reactive powers of synchronous generators 1 and 2 respectively.

# CHAPTER 5

# SIMULATIONS INVOLVING WIND POWER

In this chapter, wind power will be introduced in the system containing two synchronous generators and loads. Trends of active and reactive powers will be investigated.

## 5.1 Modelling

The block diagram which needs to be simulated is shown in Figure 5.1. The actual Simulink diagram is shown in Appendix C1.

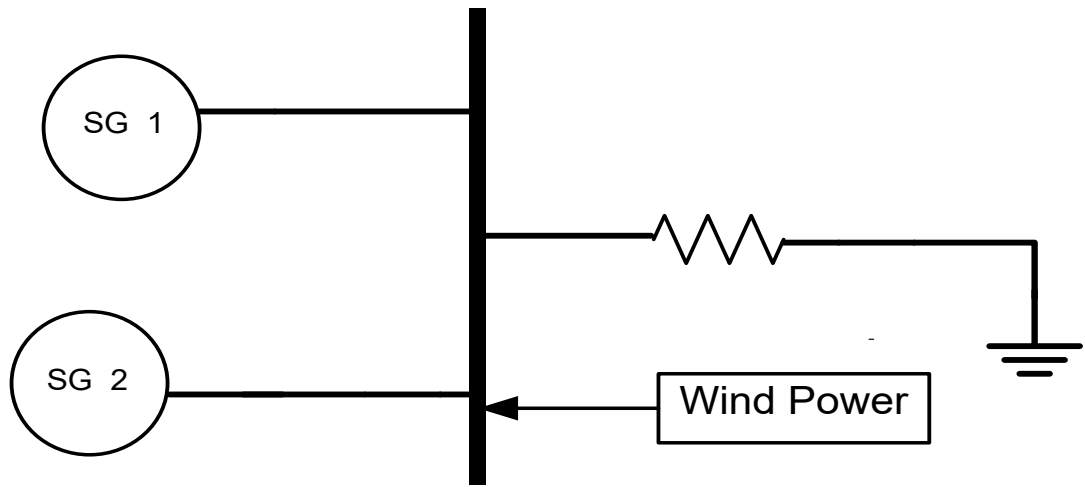

**Figure 5.1 Block diagram showing two synchronous generators with wind power and resistive load**

The scheme used to model the wind power is shown in Appendix C2. It basically introduces $I_d$ and $I_q$ into the power system. These currents translate into active and reactive power injections into the system respectively. Some basic information about types of power converters is covered in the next section.

## 5.2 Types of Power Converters in Microgrids

There are three common types of power converters used in microgrids [38]. They are called grid-feeding, grid-forming and grid-supporting power converters.

### 5.2.1 Grid-feeding Power Converters

They are the most common type of power converters used in microgrids. They can be modelled as current source in parallel with large impedance. They cannot operate in island mode unless there is a local synchronous generator which sets the voltage and frequency of the microgrid. [38]

### 5.2.2 Grid-forming Power Converters

They are modelled as ideal AC voltage sources with a known frequency and amplitude. Standby UPS is a common example of this converter. [38]

### 5.2.3 Grid-supporting Power Converters

These are the converters which are also modelled as AC voltage sources and are interfaced to the grid through impedance. Active and reactive powers of this converter depend on AC voltage of the grid and voltage source as well as the impedance through which the converter is linked to the grid. [38]

## 5.3 Relationship between Currents and Active/Reactive Powers

The active and reactive power produced by the wind is dependent on the current $I_d$ and $I_q$. In other words, higher the $I_d$, higher is the active power input into the system via wind and vice versa. Similarly, the higher the $I_q$, the higher the reactive power and vice versa.

## 5.4 Simulation Work and Results

The Simulink model for Figure 5.1 was simulated for purely resistive load. The load resistance is 121 ohms. Simulation time is 5 seconds. The solver is ode23tb. In this case, $I_d$=20 A and $I_q$= 0. Active power sharing was observed. Active powers produced by the generators and wind were equal to that absorbed by the load. Referring to Figures 5.2-5.3, active powers of SG 1 and SG 2 after 5 seconds are 0.13 MW and 0.5 MW respectively. The calculation of load power is done by using the formula:

$$P= 3 \times V_{rms} \times I_{rms} \times \cos\phi$$

Where $V_{rms}$ is the load rms phase voltage, $I_{rms}$ is the load rms phase current and $\cos\phi$ is the power factor (which is unity in this case as the load is resistive).

$V_{rms}$ and $I_{rms}$ from Figures 5.4 and 5.5 are 6353 Volts and 52.5 Amperes respectively.

Hence, load active power $P= 3 \times 6353 \times 52.5 \times 1= 1$ MW

The graphs for $V_{rms}$(phase) and $I_{rms}$(phase) are shown in Figures 5.4 and 5.5 respectively.

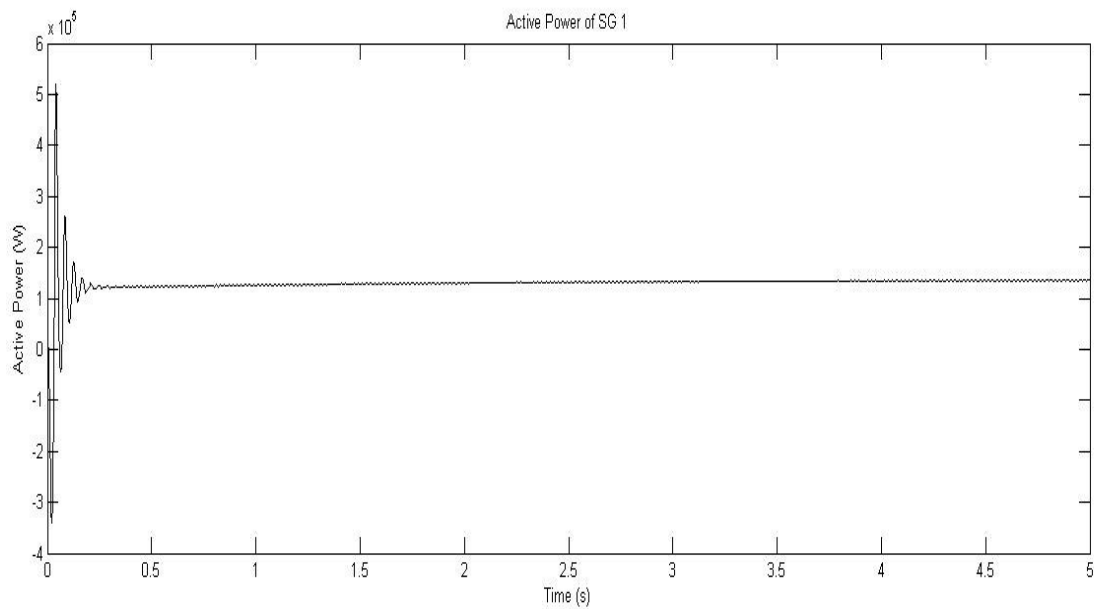

**Figure 5.2 Active Power of Synchronous Generator 1**

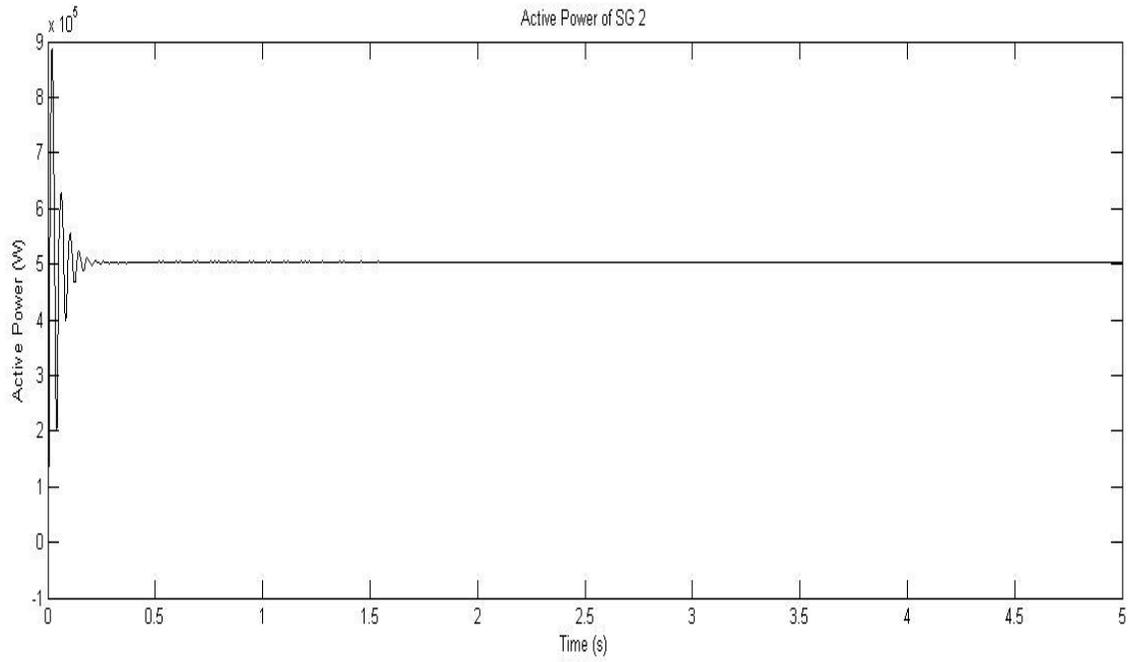

**Figure 5.3 Active Power of Synchronous Generator 2**

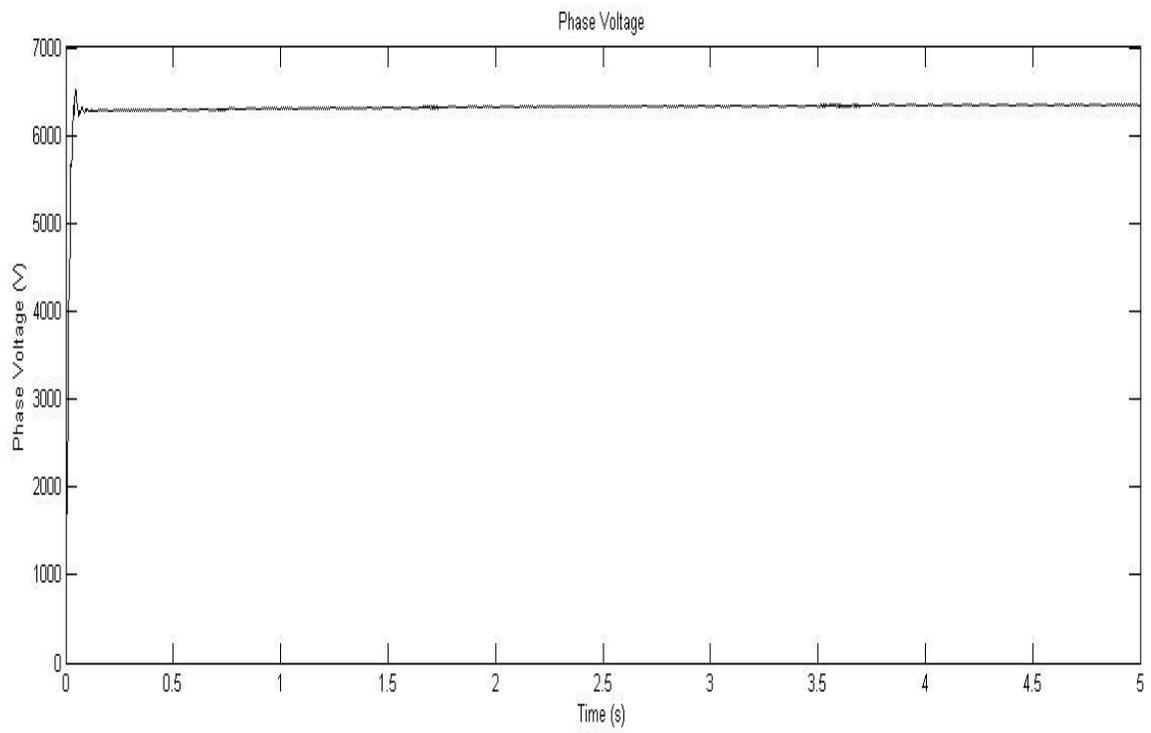

**Figure 5.4 Load RMS Phase Voltage**

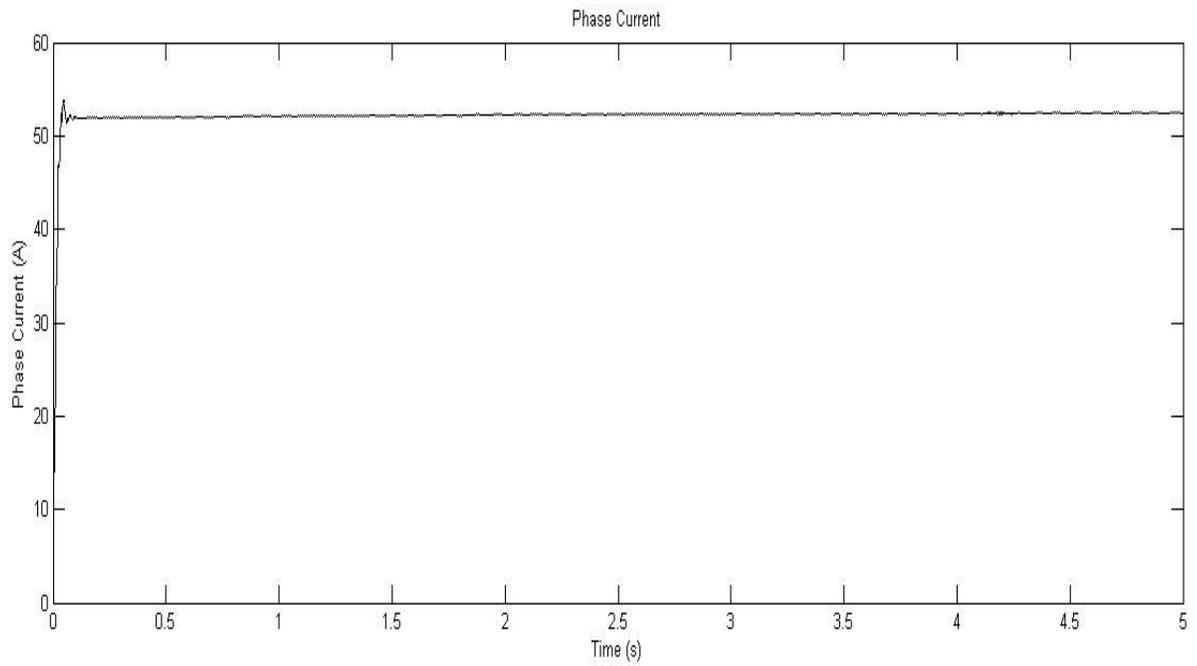

**Figure 5.5 Load RMS Phase Current**

In order to calculate the wind power, we need the rms values of voltages and currents produced by the wind power. These are shown graphically in Figures 5.6-5.7. Therefore, using the formula:

P= $3 \times V_{rms} \times I_{rms} \times \cos \phi$

Hence, Wind Active Power P= $3 \times 6350 \times 19.39$ = 0.37 MW

In short, powers produced by SG1 (0.13 MW), SG2 (0.5 MW) and Wind (0.37 MW) equal the load power (1 MW).

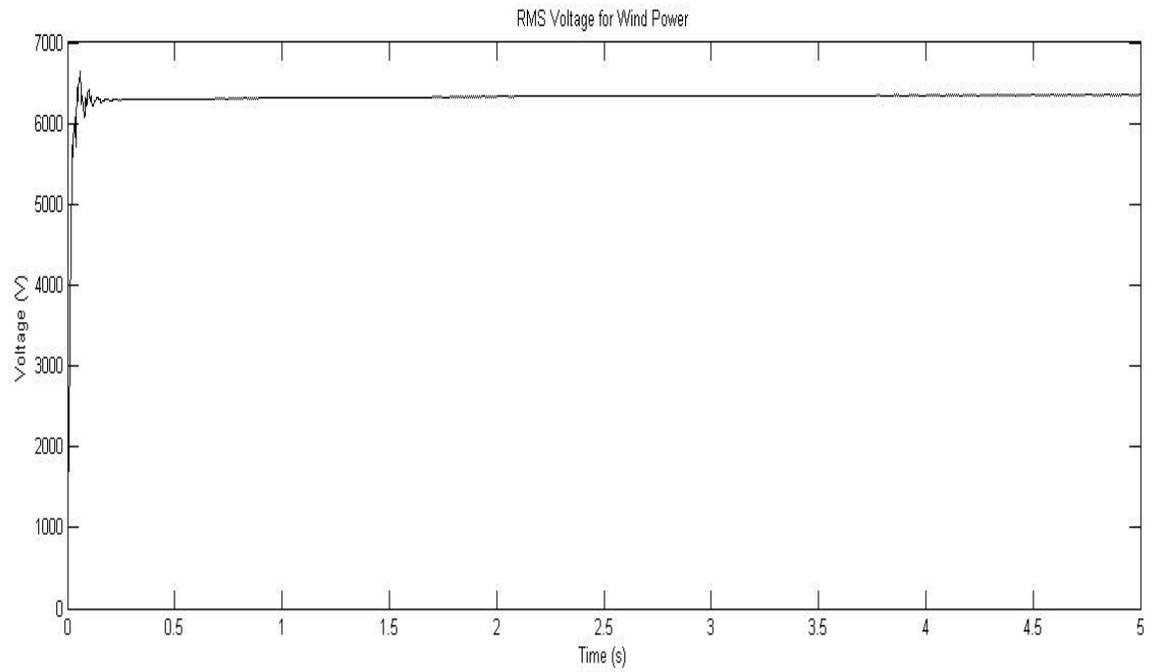

**Figure 5.6 Wind RMS Phase Voltage**

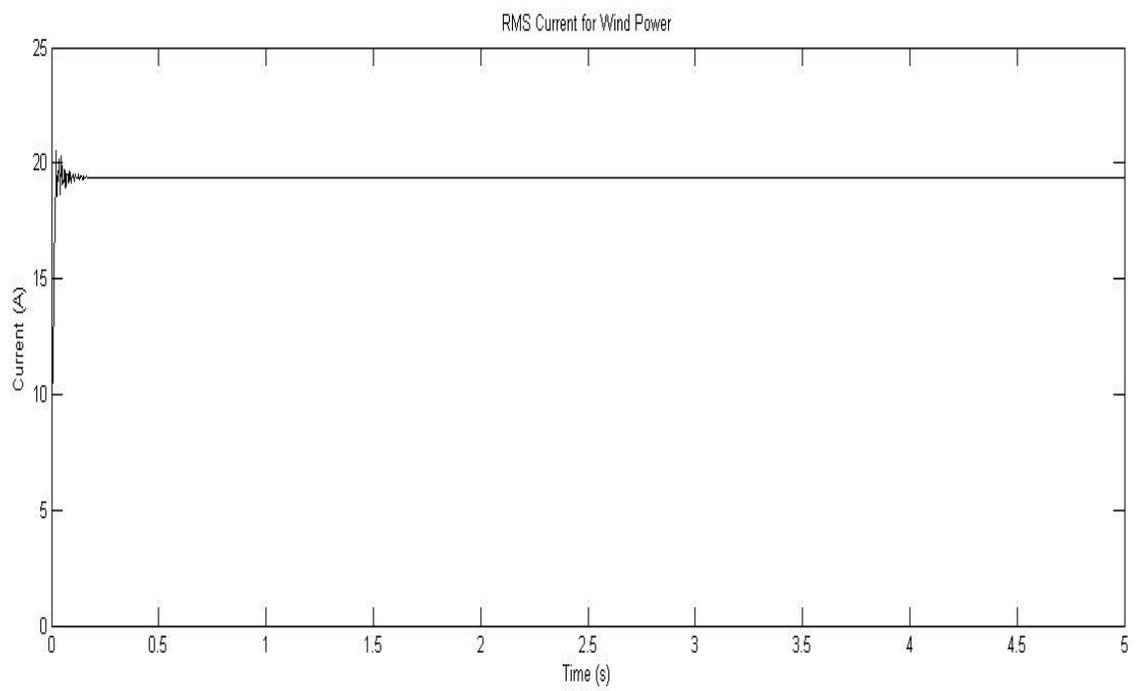

**Figure 5.7 Wind RMS Phase Current**

Now, consider the network shown in Figure 5.8. The resistive load was replaced by resistive-inductive load. Load resistance and inductance values are 121 ohms and 0.1 H. The network was simulated for 0.5 second. The solver is ode23tb.

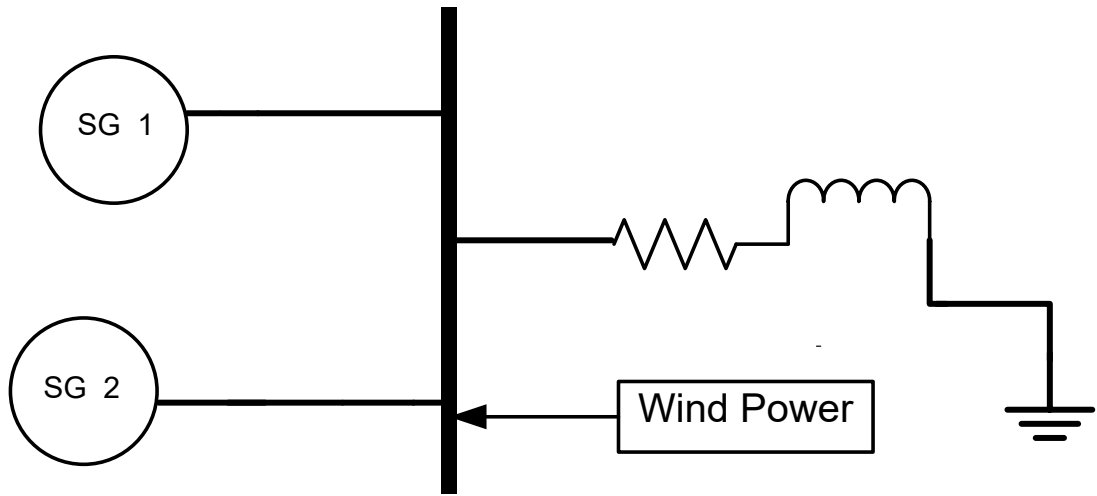

**Figure 5.8 Block diagram showing two synchronous generators with wind power and resistive-inductive load**

Now, $I_d$= 0.1A and $I_q$ =0.1 A. It was observed that as $I_q$ was increased, the reactive powers produced by the synchronous generators also increased and their sum was balanced by the reactive power absorbed by the load. This is shown in tabular form in Table 5.9

| $I_q$ (Amps) | Q1 (VARs) | Q2 (VARs) |
|---|---|---|
| 0.01 | 1.39e5 | 1.01e5 |
| 0.3 | 1.41e5 | 1.04e5 |
| 0.5 | 1.43e5 | 1.06e5 |

**Table 5.9 Reactive Power sharing for different values of $I_q$**

## 5.5 Variable Wind

In order to see the effect of variable wind on the system and how active powers are shared, $I_d$ was input as a variable wind data. $I_q$ was set to zero. The table of data used as input is shown in Appendix C2. This data is used to just observe the trend when wind speeds increase or decrease. Actual data may be ranging from 100 to 200 seconds but for simulation purposes, the data for 1 second is observed at the intervals of 0.1 second. The model of Figure 5.1 was simulated for 1 second. The solver is ode23tb. Load resistance is 121 ohms. The resulting graph for active power of SG 1 is shown in Figure 5.10. As evident from Figure 5.10, when the wind speed increases (up to 0.7 second), there is a droop in the active power of SG 1 and when the wind speed decreases (after 0.7 second), the active power of SG 1 starts to increase gradually. In other words, when there is a good availability of wind, SG 1 droops accordingly and the power requirements of the load are fulfilled by the wind and when there is low availability of wind, SG 1 provides the required load power. It must be noted that SG 2 is supplying constant power of 0.5 MW; therefore, changing wind speeds has no effect on it.

Therefore, after 1 second, SG 1 supplies 0.29 MW, SG 2 supplies 0.5 MW, wind power is 0.21 MW (3x6300x10.76) and the sum of these three powers equals the load power (1 MW).

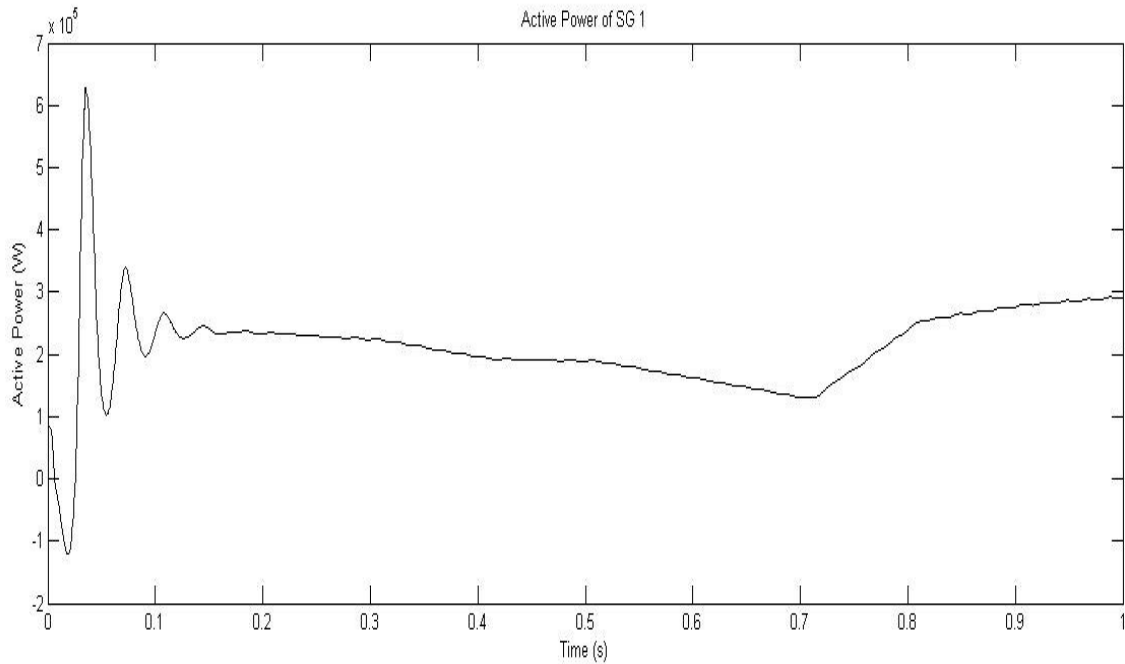

**Figure 5.10 Active Power of SG 1**

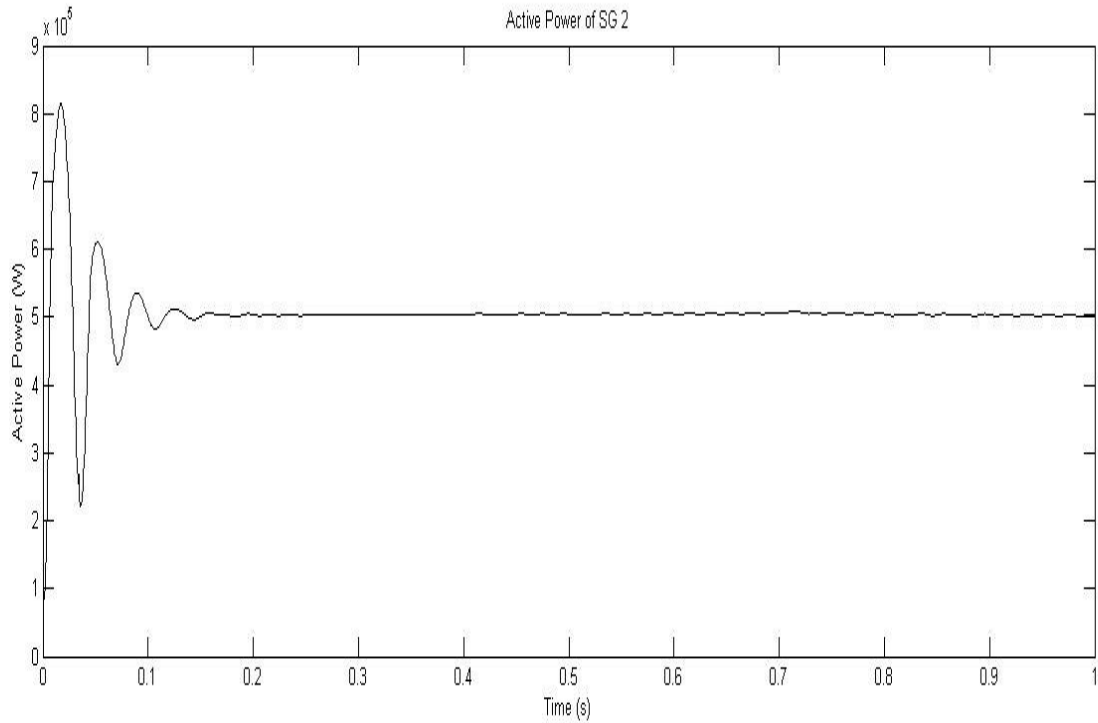

**Figure 5.11 Active Power of SG 2**

Referring to Figures 5.10-5.11, considerable amount of initial transients can be seen. One possible reason for the transients is that the generator is not doubly fed. It has a simple rotor winding which is fed by direct current. It is partially controllable as opposed to doubly fed induction machines which are fully controllable because they have two rotor windings fed with alternating current.

## 5.6 Relationship between $I_q$ and Reactive Power

In order to observe the relation between $I_q$ and reactive power of generators, $I_q$ was applied as a step input instead of a constant. The block diagram for the network is shown in Figure 5.12. Simulation time is 5 seconds and the solver is ode23tb. It was observed that as $I_q$ was doubled (from 15A to 30A), at a step time of 3 seconds, reactive power of each of the synchronous generators also doubled. This is shown graphically in Figure 5.13 and Figure 5.14. In other words, reactive power produced by each of the synchronous generators is directly proportional to the amount of reactive current injection.

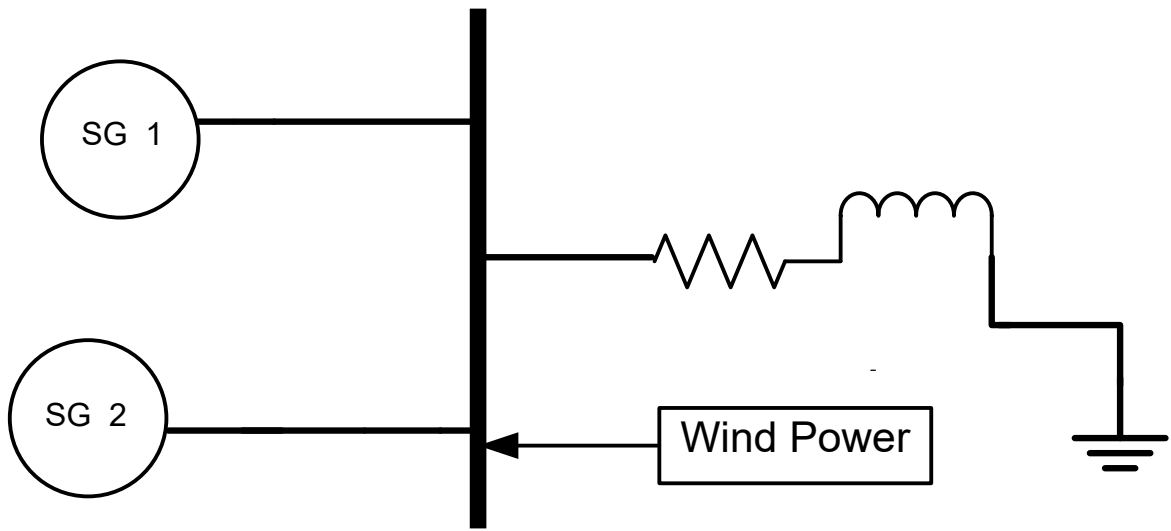

**Figure 5.12 Block diagram showing two synchronous generators with wind power and resistive-inductive load**

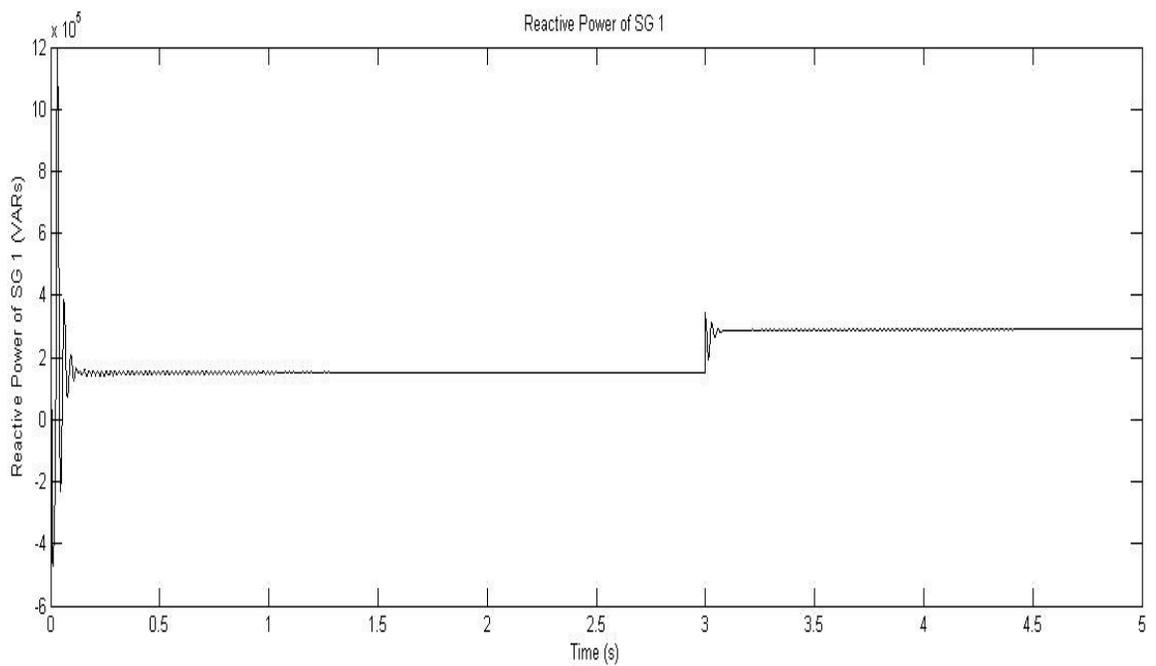

**Figure 5.13 Reactive Power of SG 1 doubles as $I_q$ is doubled**

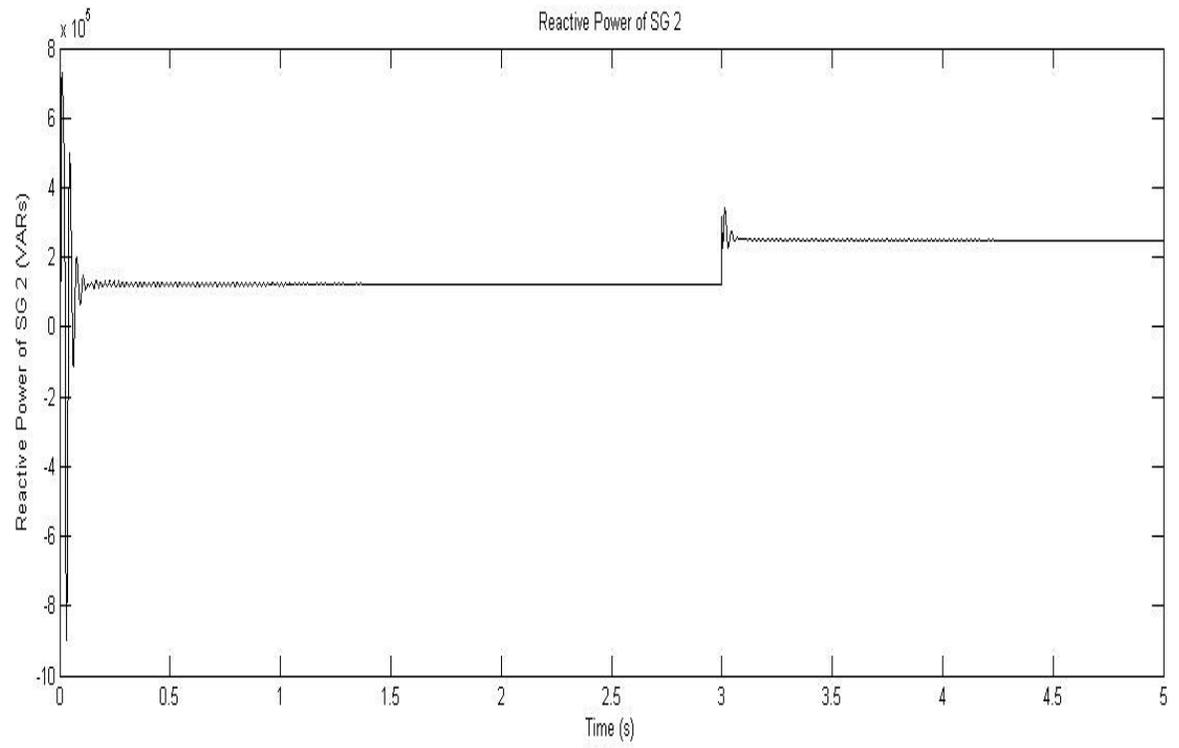

**Figure 5.14 Reactive Power of SG 2 doubles as I$_q$ is doubled**

# CHAPTER 6

# <u>FUTURE WORK AND CONCLUSIONS</u>

In this chapter, the project is concluded with some guidelines regarding future work.

## 6.1 Future Work

Some dimensions, regarding microgrids, which can be probed into, in the future are:

- In-depth investigation and evaluation of various frequency and voltage control techniques under grid-connected and grid-disconnected modes.
- Observing the transition phenomenon closely from grid-connected mode to grid-disconnected mode and vice versa. Observing the same under high penetration of wind power.
- Researching, designing and implementing protection schemes for microgrids.
- Operation of microgrids under unbalanced and non-linear loads. [24]

## 6.2 Conclusions

The project presented the major simulations leading to the microgrid implementation. Theoretical background and various development projects regarding distributed generation and microgrids were presented. Active and reactive power sharing including droop control was observed for multiple synchronous generators. The phenomenon of load transients and the concept of Master-Slave were also investigated. Moreover, the behaviour of the system was studied under wind penetration. Overall, the project was a success and nearly all the objectives were met. In short, micro-grids are an effective way of providing electric power to the consumers without disruption. This technology is relatively new and has been implemented in a few countries like USA, Japan and Canada, but in the long run, it will surely benefit all kinds of consumers and this technology is here to stay. Further know-how about this technology should be inculcated into the people by the energy sector experts and they should make them realize the importance of using this.

# References

[1]     Piagi P, Lasseter R.H, "Autonomous control of microgrids", *Power Engineering Society General Meeting, 2006. IEEE.*

[2]     Krishnamurthy.S, Jahns T.M, Lasseter R.H., "The operation of diesel gensets in a CERTS microgrid", *Power and Energy Society General Meeting - Conversion and Delivery of Electrical Energy in the 21st Century, 2008 IEEE.*

[3]     Ling Su, Jianhua Zhang, Weishi Miao, Qi Yang, Ruoxi Liu, "Study on Control Strategy for Islanded Microgrid Based on Microturbine", *Power and Energy Engineering Conference (APPEEC), 2010 Asia-Pacific.*

[4]     Yong Xue, Jiamei Deng, Shuangbao Ma, "Power flow control of a distributed generation unit in micro-grid", *Power Electronics and Motion Control Conference, 2009. IPEMC '09. IEEE 6th International.*

[5]     Jia Yaoqin, Liu Dingkun, Pan Shengkui, "Improved droop control of parallel inverter system in standalone microgrid", *Power Electronics and ECCE Asia (ICPE & ECCE), 2011 IEEE.*

[6]     Wang Yang, Ai Xin, Gao Yang, "Microgrid's operation-management containing distributed generation system", *Electric Utility Deregulation and Restructuring and Power Technologies (DRPT), 2011 4th International Conference.*

[7]     Tang Yongjun, Liu Dong, Ruan Qiantu, "Optimal Allocation of Distributed Generation and Its Parallel Computation Considering Energy-saving Dispatching", *Automation of Electric Power System, vol. 32, No. 7, pp. 92-98 , 2008.*

[8]     Rowe C, Summers T.J, Betz, R.E., "A novel parallel voltage and current control scheme implementing P-F and Q-V droop in a microgrid", *Power Electronics and Applications (EPE 2011), Proceedings of the 2011-14th European Conference.*

[9]     Basak, P, Saha, A.K., Chowdhury, S., Chowdhury, S.P., "Microgrid: Control techniques and modeling", *Universities Power Engineering Conference (UPEC), 2009 Proceedings of the 44th International.*

[10]    Zhang Jie, Wu Peng, Hong Jie, "Control strategy of microgrid inverter operation in Grid-connected and Grid-disconnected modes", *Electric Information and Control Engineering (ICEICE), 2011 International Conference.*

[11]    Wei Huang, Miao Lu, Li Zhang, "Survey on microgrid control strategies", *ICSGCE 2011: 27–30 September 2011, Chengdu, China.*

[12]    Huang Jiayi, Jiang Chuanwen, Xu Rong, "A review on distributed energy resources and microgrid", *Renewable and Sustainable Energy Reviews 12 (2008) 2472-2483.*

[13]    M.K. Donnelly, J.E. Dagle, D.J. Trudnowski, and G.J. Rogers, "Impacts of the distributed utility on transmission system stability*," IEEE Transactions on Power Systems, vol. 11, Ĉ. 2, pp. 741-746, May 1996.*

[14]    Zhang Guoju, Tang Xisheng, and Qi Zhiping, "Application of Hybrid Energy Storage System of Super-capacitors and Batteries in a Microgrid", *Automation of Electric Power System, vol. 34, No. 12, pp. 85-90, 2010.*

[15]    CERTS, "Integration of distributed energy resources – The CERTS microgrid concept*", submitted to the U.S Department of Energy April 2002.*

[16]    R. H. Lasseter, "Microgrid" *Power Engineering Society Winter Meeting, vol. 1, pp. 146–149, 2001.*

[17]    W.El-Khattam, M.A.Salama, "Distributed generation technologies, definitions and benefits", *Electric Power Systems Research Volume 71, Issue 2, October 2004, Pages 119–128.*

[18]    G. Pepermans, J. Driesen, D. Haeseldonckx, R. Belmans, W. D'haeseleer, "Distributed generation: definition, benefits and issues", *Energy Policy, Volume 33, issue 6, April 2005, Pages 787-798.*

[19]    A.A. Bayod Rújula, J. Mur Amada, J.L.  Bernal-Agustín, J.M. Yusta Loyo, J.A, Domínguez Navarro, "Definitions for Distributed Generation: a revision".

[20]    "Assessment of Distributed Generation Technology Applications" - A report prepared by the Resource Dynamics Corporation for the Maine Public Utilities Commission, February 2001, <www.distributed-generation.com/Library/Maine.pdf>, (Accessed :22 July, 2012).

[21]    Hans B.Puttgen, Paul R. Macgregor, Frank C.Lambert, "Distributed Generation: Semantic Hype or the Dawn of a New Era", *IEEE Power and Energy Magazine, January/February 2003, Volume 1, Issue 1.*

[22]    Nigim, K.A., Wei-Jen Lee, "Micro Grid Integration Opportunities and Challenges", *Power Engineering Society General Meeting, 2007. IEEE.*

[23]    Sara Stroud, "Microgrids get big", 2 March, 2011, <http://sustainableindustries.com/articles/2011/02/microgrids-get-big>, (Accessed: 24 August, 2012).

[24]    A.A. Salam, A.Mohamed, M.A.Hannan, "Technical Challenges on Microgrids", *ARPN Journal of Engineering and Applied Sciences, Volume 3, No.6, December 2008.*

[25]    Shahab Ahmad Khan, Reshadat Ali, Sharib Hussian, "Introduction to Micro-grid",

<http://www.slideshare.net/Shahabkhan/microgrid-presentation>, (Accessed: 22 July, 2012).

[26]    Siemens White Paper, "Microgrids", <http://www.energy.siemens.com/fi/pool/hq/energy-topics/smart-grid/documents/3558_White%20paper%20Microgrids_EN_LR.pdf>, (Accessed: 22 July, 2012).

[27]    Roger G.Lawrence, "Distributed Generation", September 2003, <http://www.rglsolutions.com/Distributed_Generation.htm>, (Accessed: 22 July, 2012).

[28]    Johnzactruba, Lamar Stonecypher, "A System of Systems - Microgrids Poised to Soar in Popularity", 19 May, 2011, <http://www.brighthub.com/engineering/electrical/articles/90436.aspx>, (Accessed: 22 July, 2012).

[29]    Product Documentation, "SimPower Systems", <http://www.mathworks.co.uk/help/toolbox/physmod/powersys/ug/bs_fxkn.html>, (Accessed: 08 August, 2012).

[30]    Product Documentation, "Simulink", <http://www.mathworks.co.uk/help/toolbox/simulink/ug/f11-69449.html>, (Accessed: 08 August, 2012).

[31]    "Parallel Operation of Synchronous Generators", <http://www.csi.uottawa.ca/~rhabash/ParSynGen>, (Accessed: 24 July, 2012).

[32]    Andrew Mark Bollman, "An Experimental Study of Frequency Droop Control in a low inertia Microgrid", *M.S. Thesis, Department of Electrical & Computer Engineering, Graduate College of the University of Illinois at Urbana-Champaign, 2009.*

[33]    Meghdad Fazeli, "Wind Generator-Energy Storage Control Schemes for Autonomous Grid", *Ph.D. Dissertation, Department of Electrical & Electronic Engineering, University of Nottingham, U.K., 2010.*

[34]    Stephen J. Chapman, "Synchronous Generators", in *Electric Machinery Fundamentals, 4th Edition, McGraw -Hill, Page 304.*

[35]    Meena Agarwal, Arvind Mittal, "Microgrid technological activities across the globe: A Review", May 2011, <www.arpapress.com/Volumes/Vol7Issue2/IJRRAS_7_2_07.pdf>, (Accessed: 22 July, 2012).

[36]    Heath Reidy, "China to build first direct current micro-grid", 20 March, 2012,

<http://www.renewable-energy-technology.net/grid-storage/china-build-first-direct-current-microgrid>, (Accessed: 23 July, 2012).

[37]    Vaclovas Miskinis, Egidijus Norvaisa, Arvydas Galinis, Inga Konstantinaviciute, "Trends of distributed generation development in Lithuania", *Energy Policy, Volume 39, Issue 8, August 2011, Pages 4656-466.*

[38]    Joan Rocabert, Alvaro Luna, Frede Blaabjerg, Pedro Rodriguez, "Control of Power Converters in AC Microgrids*", IEEE Transactions on Power Electronics, Volume 27, Issue 11, May 2012, Pages 4734-4749.*

# Appendix A1

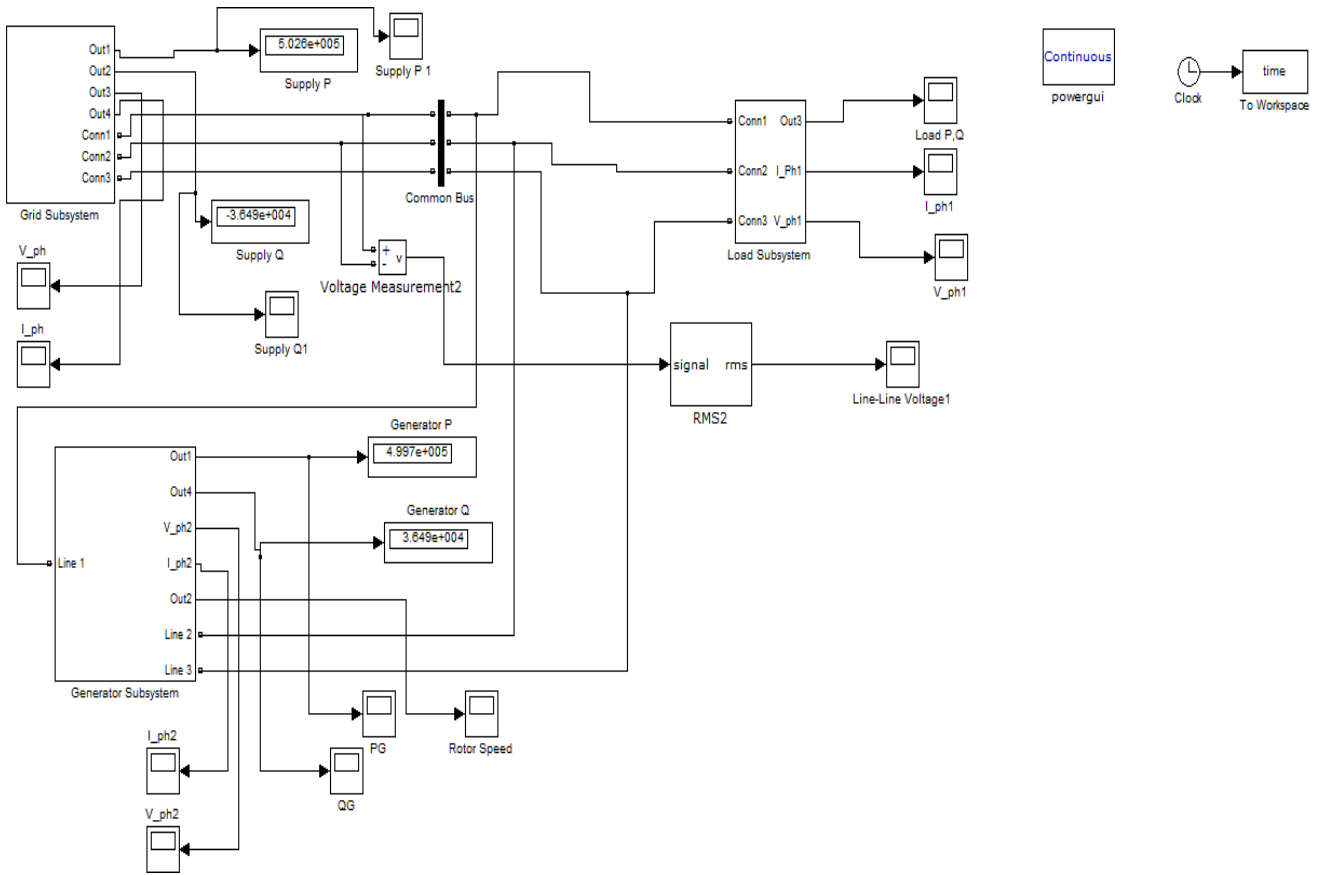

**Simulink diagram for grid, synchronous generator and resistive load**

# Appendix A1 (Continued)

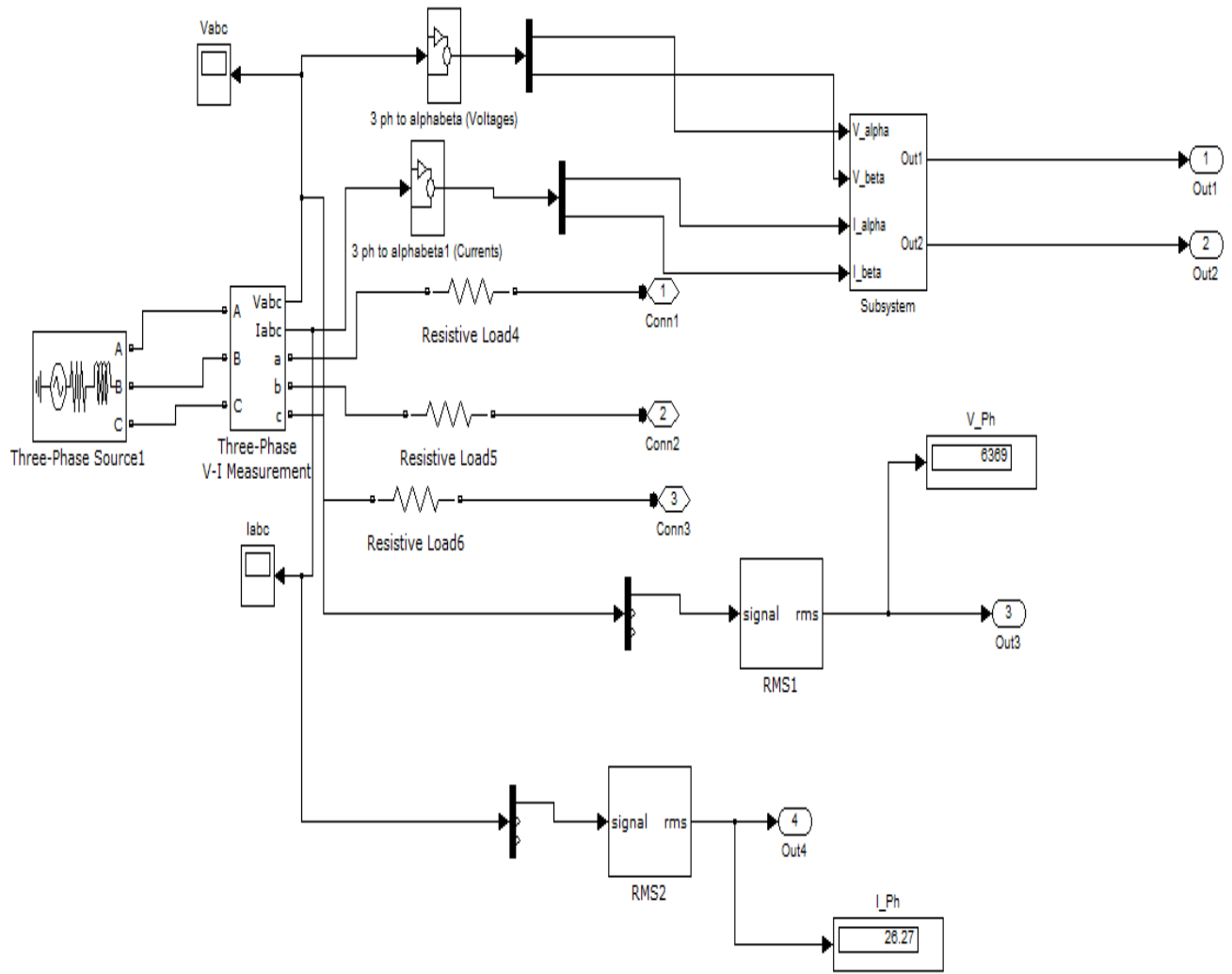

**Simulink diagram for grid subsystem**

# Appendix A1 (Continued)

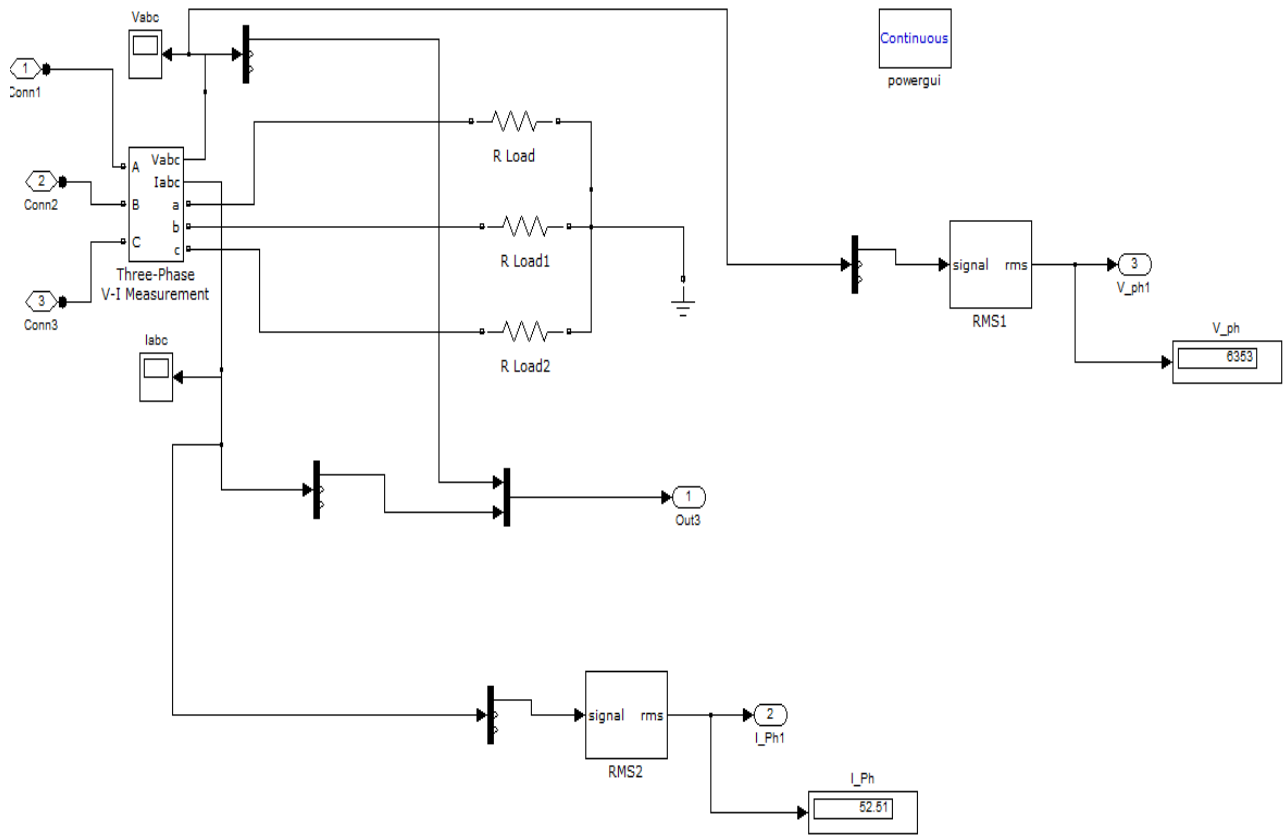

**Simulink diagram for load subsystem**

# Appendix A1 (Continued)

**Simulink diagram for synchronous generator subsystem**

# Appendix A2

As discussed in Section 3.2.5 that load angle takes a very long time to reach a steady state value. So, the model consisting of grid, synchronous generator and resistive load was simulated for 500 seconds and load angle was plotted. The plot is shown below. From this plot, the steady state value of load angle $\delta$ is 22.5 degrees.

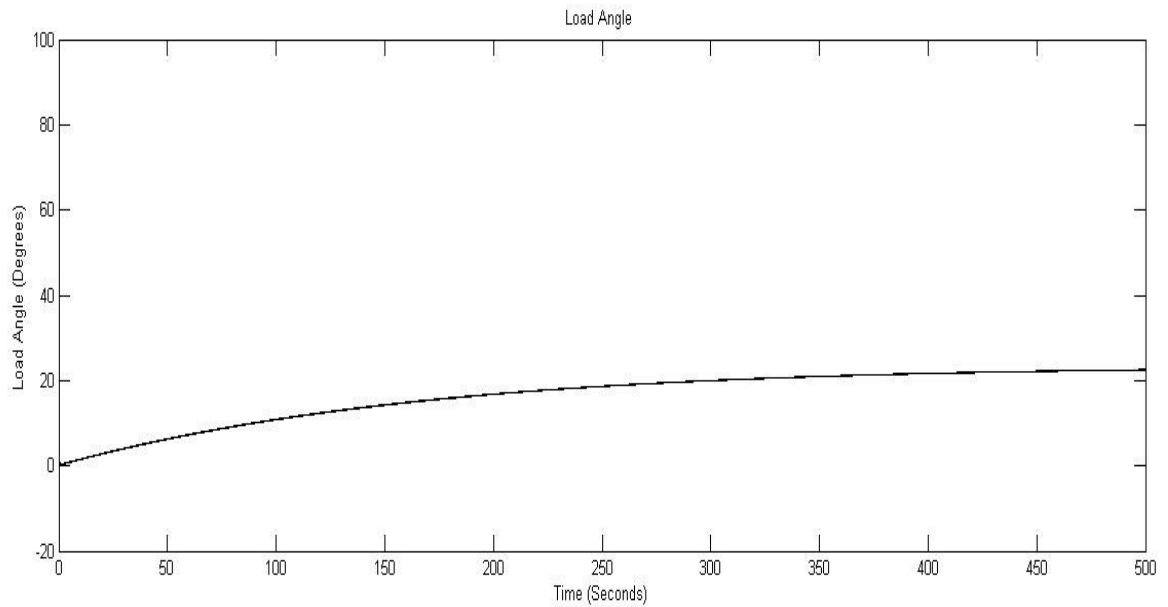

**<u>Figure showing graphical representation for load angle</u>**

Now, to confirm, that the output power of synchronous generator is actually 0.5 MW, let us turn our attention to the well known formula:

$$P = 3\,\frac{VtEf}{Xs}\,\sin\delta \quad \ldots\ldots\ldots\ldots\ldots\ldots\ldots.. \ (2.1)$$

# Appendix A2 (Continued)

In equation 2.1, $E_f = \omega_e \times$ flux$= 314 \times 28 = 8.7$ KV. But as we want to use rms convention (SimPower works on peak convention), it has to be divided by $\sqrt{2}$. Hence, $E_f = 6.2$ KV. $V_t = 6.3$ KV. $X_S$ is the ratio of $E_f$ and peak $I_{sc}$ (short circuit current). The ratio is calculated as:

$X_s = \dfrac{Ef}{Isc} = 6200/70 = 88.6$ Ohms. Peak $I_{sc}$ was calculated by short circuiting the stator terminals and measuring the current at the rated speed of 1500 RPM (157 radians/sec). Load angle $\delta$ is 22.5 degrees. Hence, putting these values in equation (2.1) yields generator output power equal to 0.5 MW.

# Appendix A3

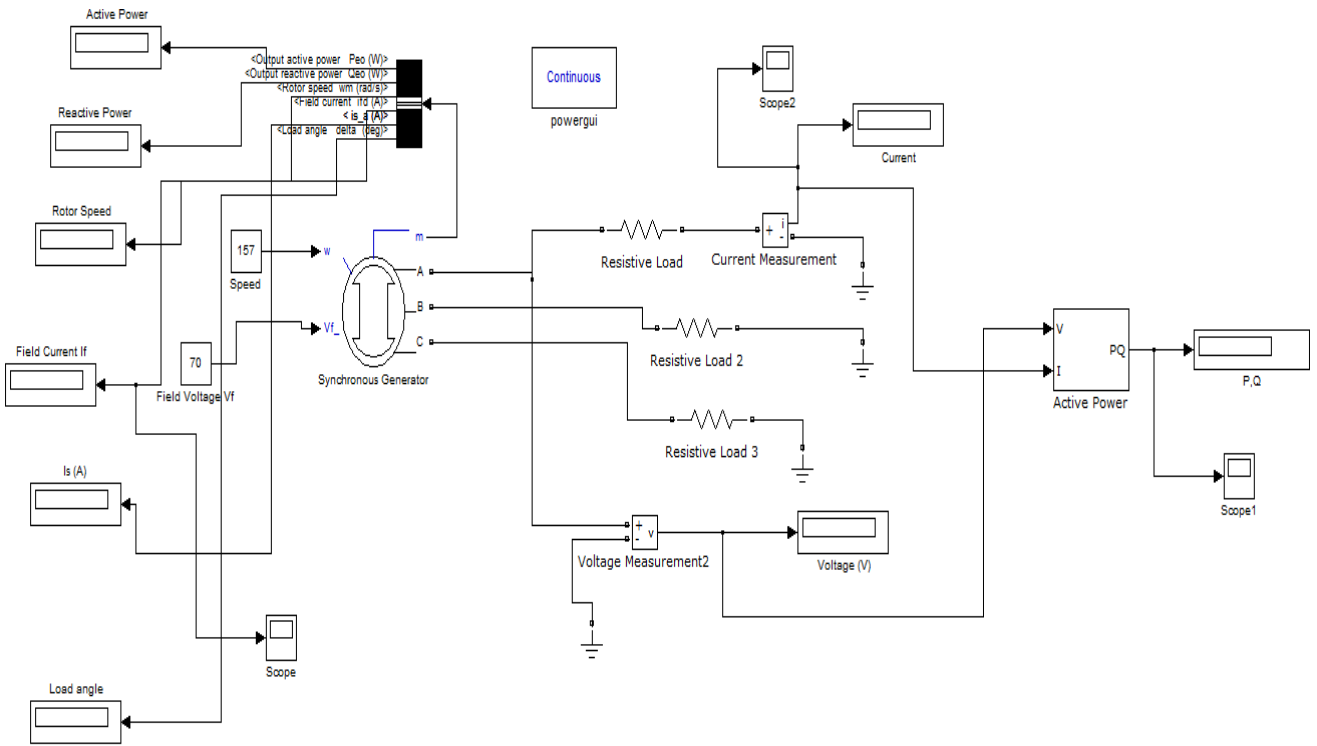

**Simulink diagram for trends in load angle (Synchronous generator, Resistive load)**

# Appendix A4

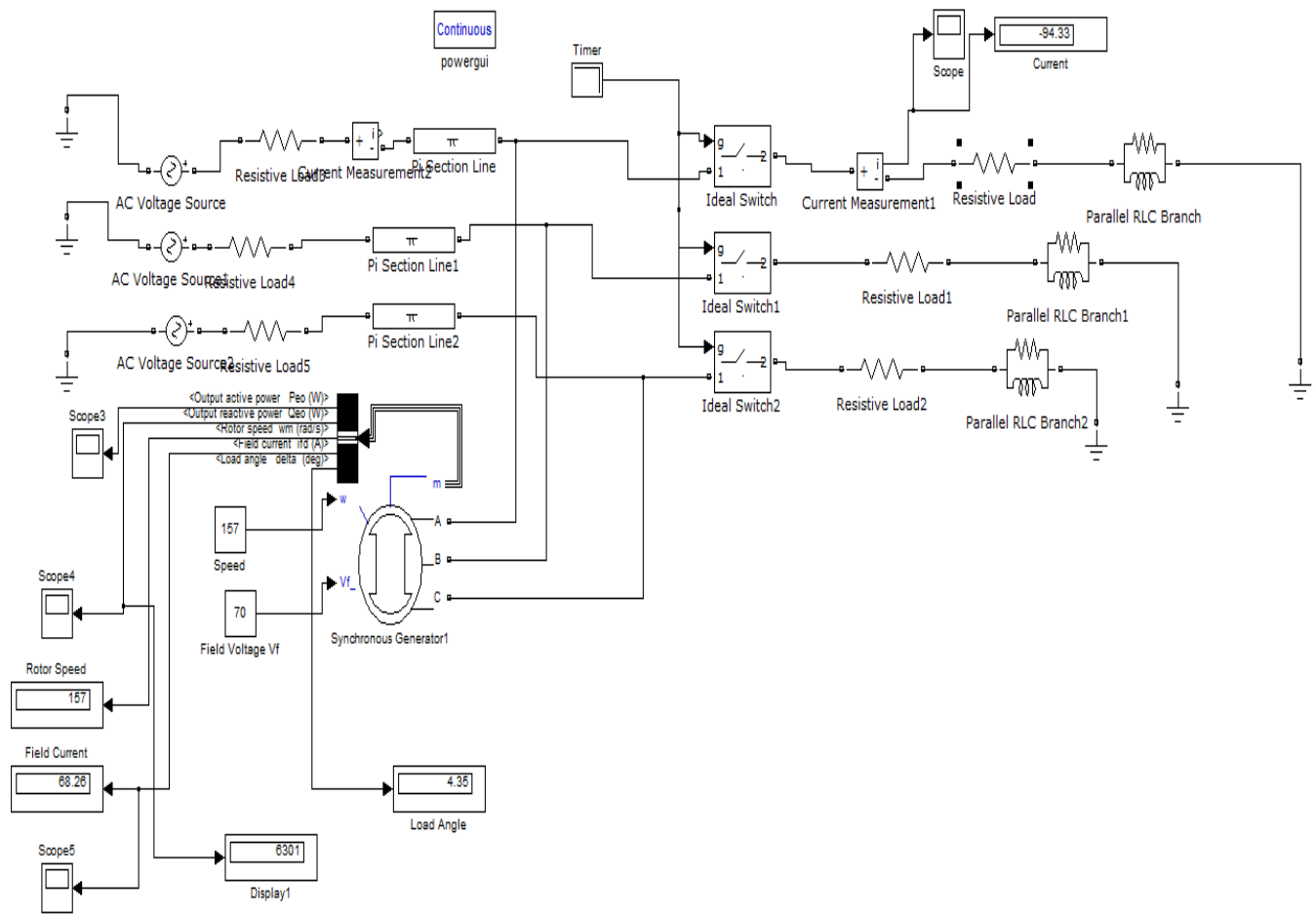

**Simulink diagram for trends in load angle (Synchronous generator, grid and RL Load)**

# Appendix A5

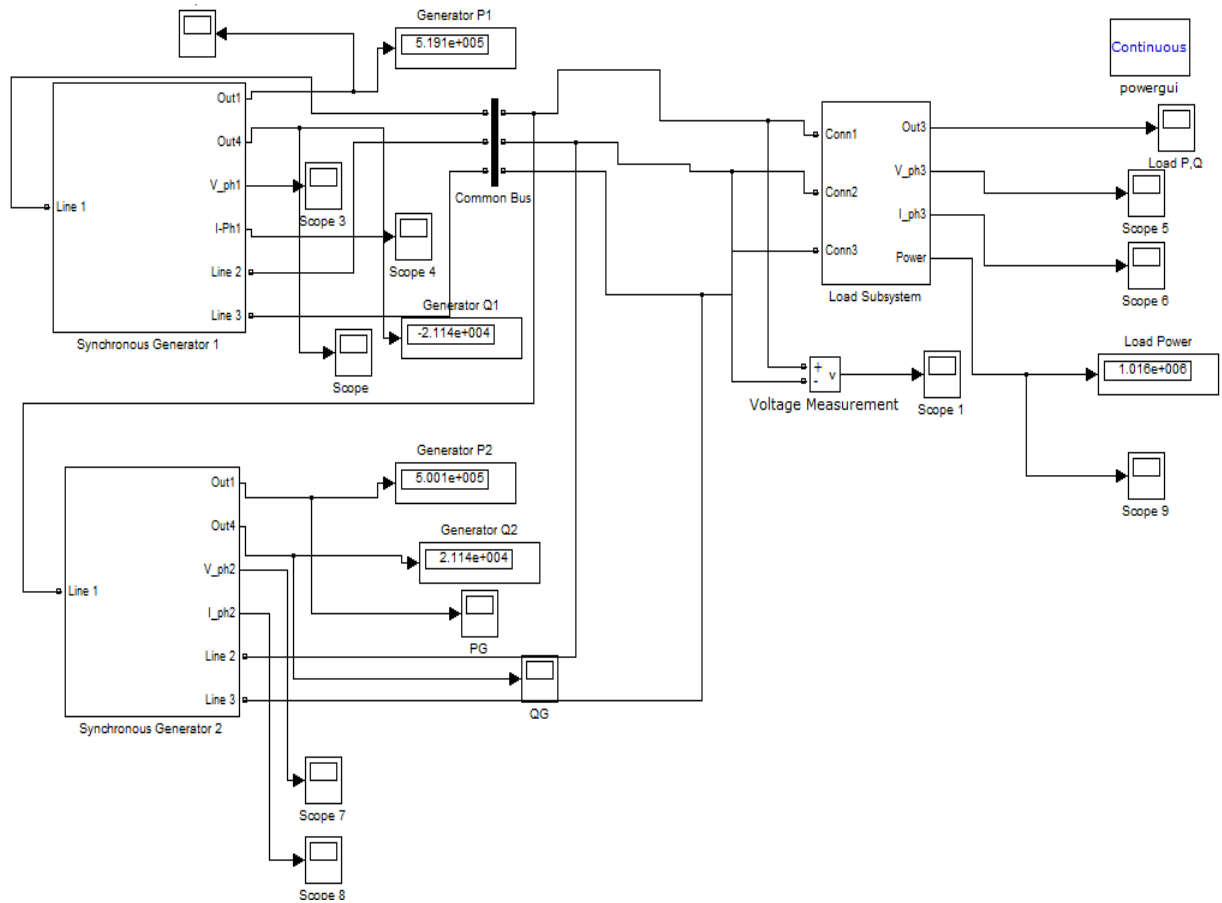

**Simulink diagram for two synchronous generators and loads**

(Subsystems for Synchronous generators and loads are the same as given in Appendix A1)

# Appendix B1

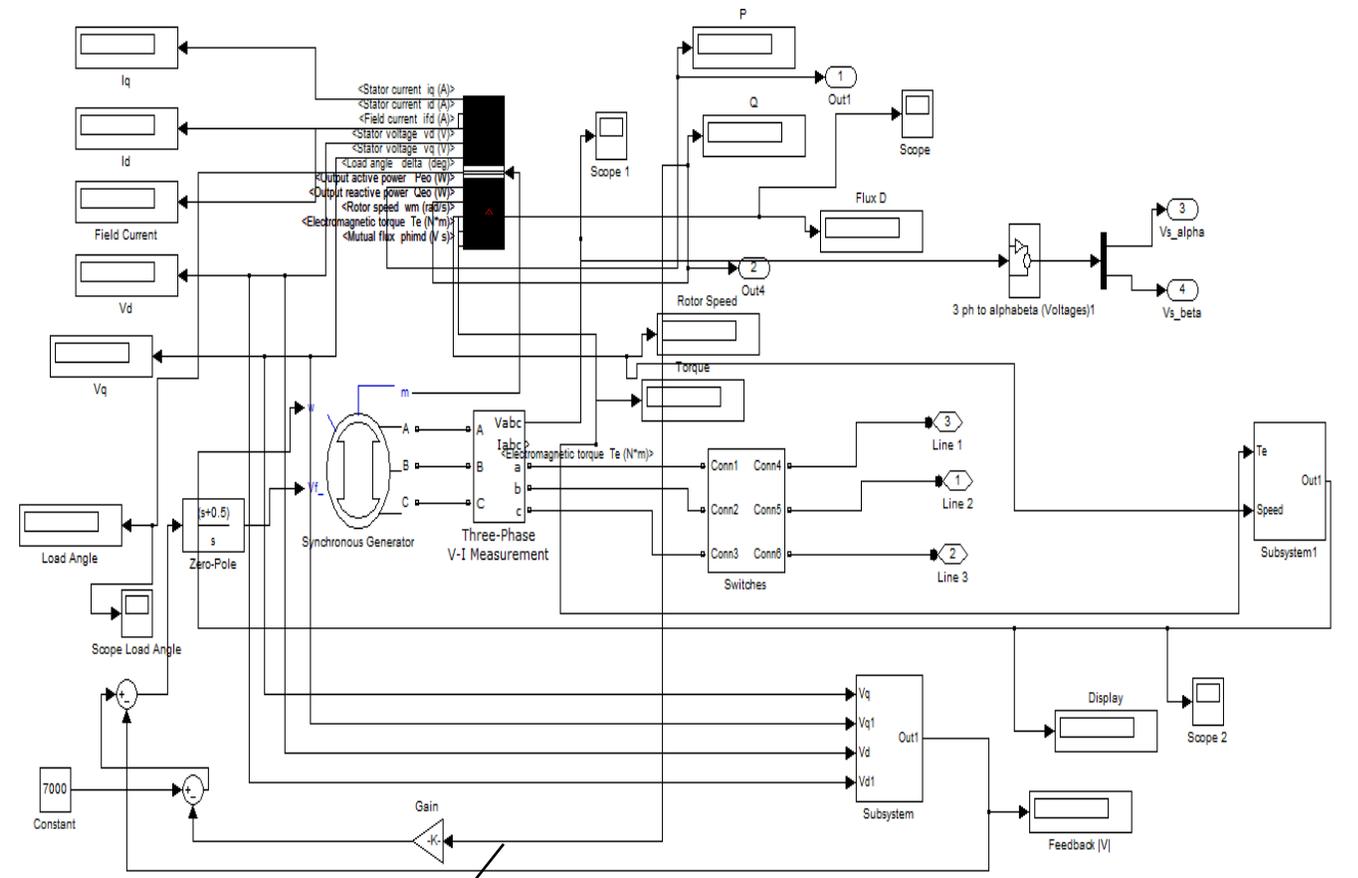

**Simulink diagram for reactive power-voltage droop control of synchronous generators**

**Control Scheme for implementing Reactive Power-Voltage droop**

# Appendix B2

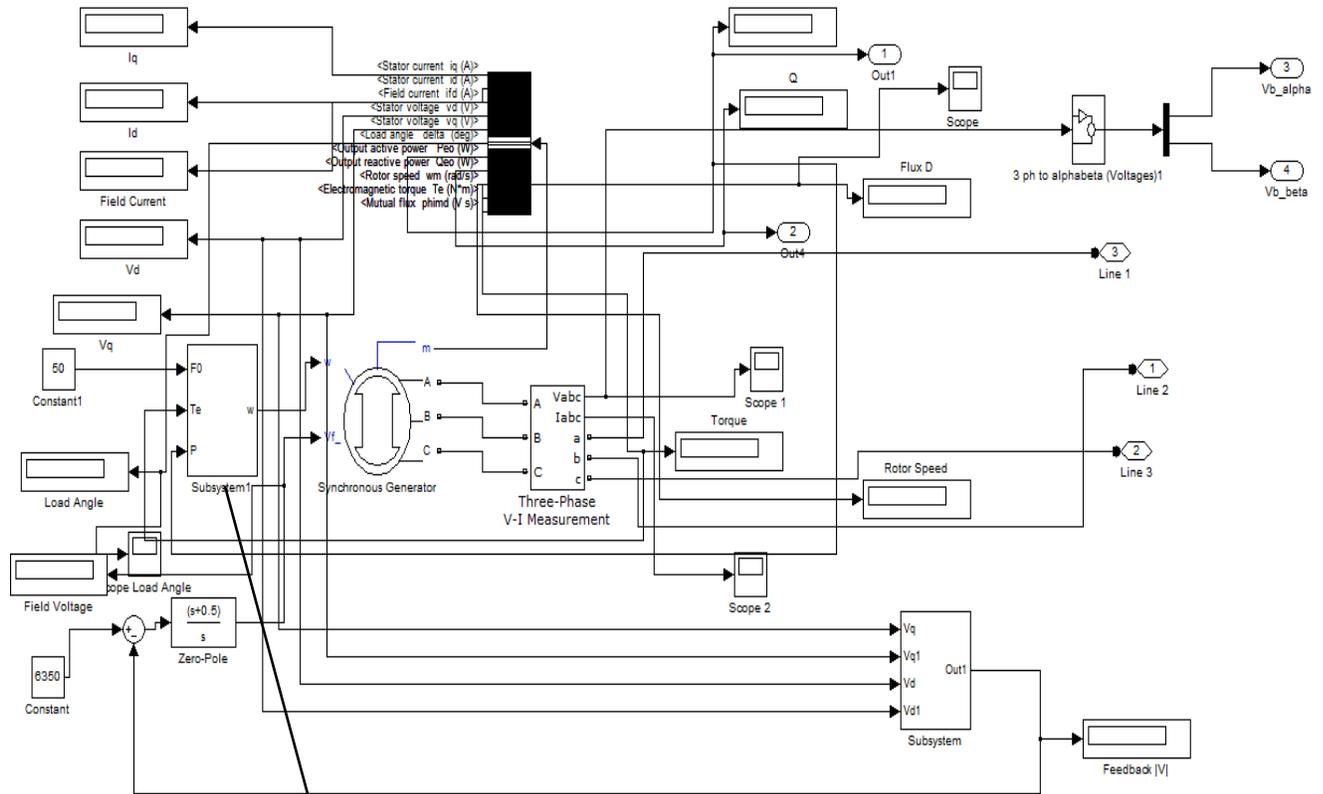

**Simulink diagram for active power-frequency droop control of synchronous generators**

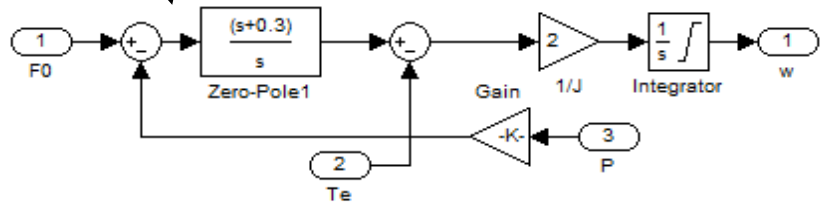

**Control Scheme for implementing Active Power-Frequency droop**

# Appendix C1

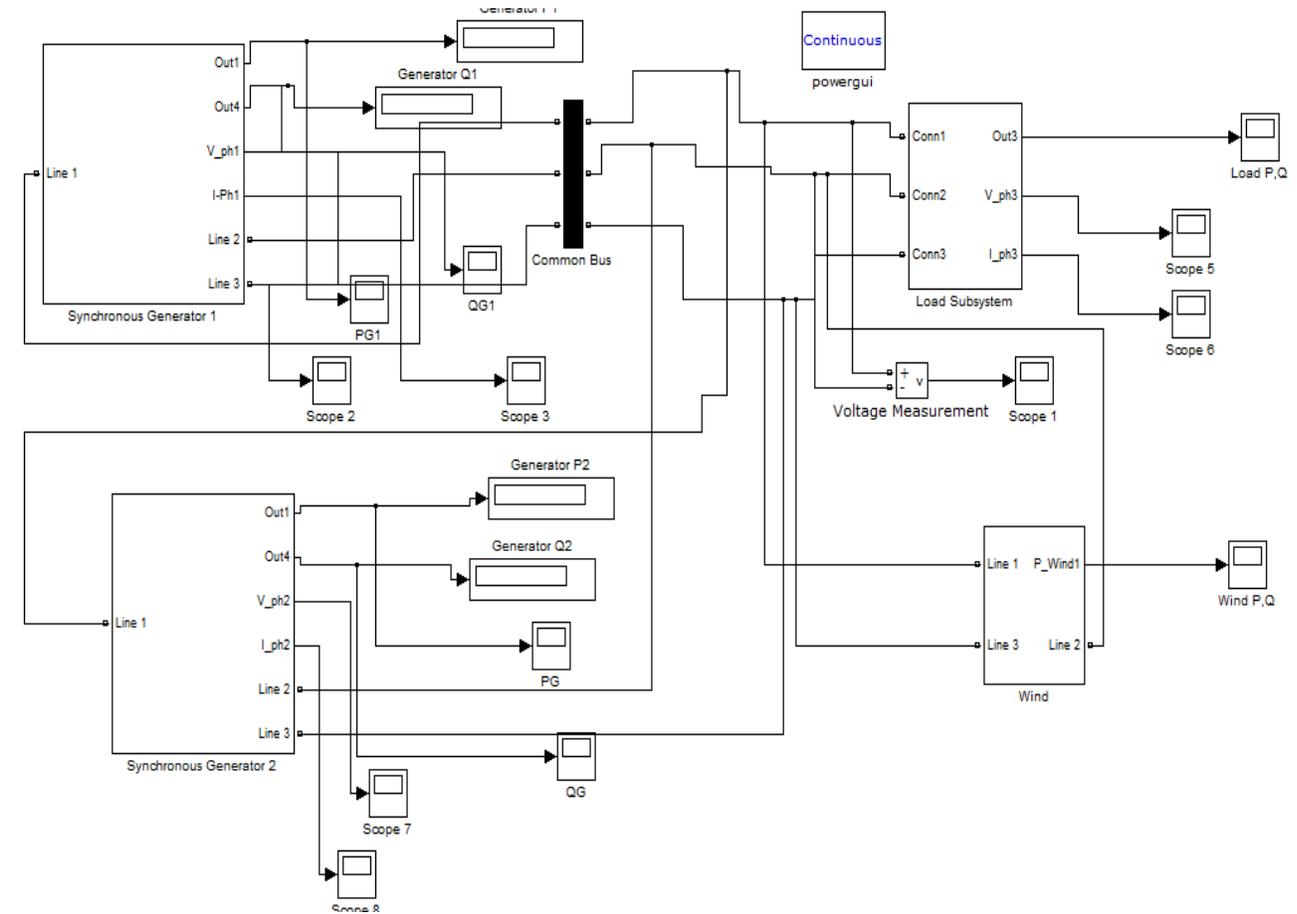

**Simulink diagram for the system consisting of synchronous generators, loads and wind power**

# Appendix C2

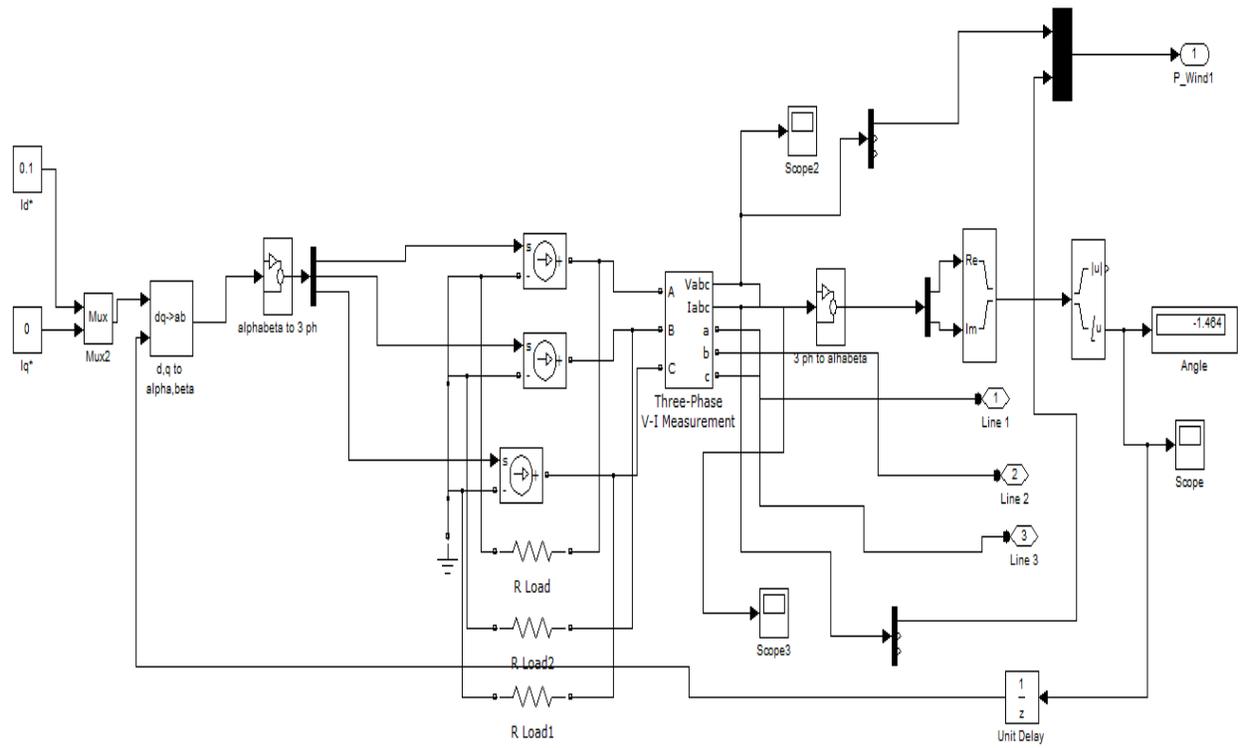

**Simulink diagram for modelling the wind power**

| Time (s) | Wind Speed (m/s) |
|----------|------------------|
| 0.0 | 13.2 |
| 0.1 | 13.9 |
| 0.2 | 14.2 |
| 0.3 | 14.8 |
| 0.4 | 16.5 |
| 0.5 | 16.7 |
| 0.6 | 18.4 |
| 0.7 | 20.1 |
| 0.8 | 13.2 |
| 0.9 | 11.9 |
| 1.0 | 11.2 |

**Table of data for implementation of variable wind**